%% file: main.tex
\documentclass[10pt,a4paper]{article}

\usepackage{indentfirst} 
\usepackage{float}
\usepackage[margin=1.25in]{geometry}
\usepackage{placeins}
\usepackage{comment}
\usepackage[utf8]{inputenc}
\usepackage[italian=guillemets]{csquotes}
\usepackage[english]{babel}
\usepackage[style=authoryear]{biblatex}
\usepackage{amsmath}
\usepackage{amsfonts}
\usepackage{amssymb}
\usepackage{graphicx}
\usepackage{lmodern}
\usepackage{hyperref}
\usepackage{braket}
\usepackage{tikz}
\usepackage[T1]{fontenc}
\usepackage{enumitem}
\usepackage{amsmath}
\usepackage{algorithm}
\usepackage{algpseudocode}
\usepackage[toc,page]{appendix}
\usepackage{subcaption}
\usepackage{appendix}
\usepackage{todonotes}
\usepackage{booktabs}
\usepackage{multirow}
\usepackage[normalem]{ulem}
\usetikzlibrary{decorations.pathreplacing}
\definecolor{darkgreen}{RGB}{100,200,0}
\usepackage{xcolor}
\usepackage{dsfont}
\usepackage{siunitx}
\usepackage{wrapfig}

\newcommand{\tilS}{\Tilde{S}_t}
\newcommand{\tilSminus}{\Tilde{S}_{t-1}}

\input{preamble}

\usepackage{tabularx}

\begin{document}

\title{Deep reinforcement learning for optimal trading with partial information}
\author{Andrea Macr\`i$\,^{1}$, Sebastian Jaimungal$\,^{2, 3}$ and Fabrizio Lillo$^{\,1,4}$ \\
\\                              
\small{$^1$ Scuola Normale Superiore, Pisa}\\
\small{$^2$ Department of Statistical Sciences, University of Toronto}\\
\small{$^3$ Oxford-Man Institute for Quantitative Finance, University of Oxford}\\
\small{$^4$ Dipartimento di Matematica, University of Bologna}}
\date{\today}
\maketitle
\begin{abstract}
    {Reinforcement Learning (RL) has attracted increasing interest in financial applications, including optimal trading and execution. However, the use of RL for optimal trading strategies that exploit latent information in the market has been, to the best of our knowledge, subject to little attention. In this paper, we consider an optimal trading problem in which the trading signal follows an Ornstein-Uhlenbeck process with regime switching parameters. The problem is naturally formulated as a partially observable Markov decision problem, requiring a trader to infer latent information directly from the history of the observable process. We combine recurrent neural networks (RNN) with RL to address both the filtering and trading components of the problem. More specifically, we propose three distinct RL-based approaches, on top of which we incorporate RNN to filter the latent state of the environment where the agent is trading. The first, a one-step approach, directly encodes hidden states from the GRU into the RL trader. The second and third are two-step methods: one that uses posterior regime probability estimates for the mean reverting regimes, while the other relies on forecasts of the next signal value. Through extensive simulations with increasingly complex Markovian regime dynamics for the trading signal's parameters, as well as an empirical application to equity pairs trading, we find that feeding posterior probability estimates on the latent long-run mean reversion regime achieves superior cumulative rewards and exhibits more interpretable strategies, in contrast with the other two approaches. Overall, our results show that the quality and structure of the information supplied to the agent are important for trading performance and policy interpretability. }    
\end{abstract}

\section{Introduction}
Optimal trading strategies have been the subject of an active and flourishing area of research since the advent of electronic markets several decades ago. The development of more efficient quoting systems and the faster spread of information have led major market participants, on both sell and buy sides, to embed ever more granular information into their trading algorithms. 

In model-based approaches to algorithmic trading, recent work includes the incorporation of trading signals into the classical optimal stochastic control formulation of trading, starting with the works of \cite{garleanu2013dynamic} in discrete time and \cite{cartea2016incorporating} and \cite{casgrain2019trading} (with latent signals) in continuous time.
Thanks to the advent of machine learning (ML), research has begun to focus on model-agnostic approaches to incorporate trading signals. In this regard, among the many methods proposed in literature, those that propose forecasting algorithms that  use  Recurrent Neural Networks (RNNs) show the greatest promise. For example, \cite{tsantekidis2020using}, investigate many architectures for  predicting future price levels starting from Limit Order Book  (LOB) information; \cite{sirignano2021universal} where the authors use 
high frequency data containing all orders, transactions and order cancellations for approximately 1000 stocks traded on NASDAQ to predict their returns; \cite{zhang2019deeplob} propose a new convolutional neural network on top of an LSTM layer to forecast stock returns at high frequency and finally \cite{kolm2023deep} where the authors compare the performance of several machine learning architectures in extracting excess return information for multiple horizons. Extracting market signals, however, is only one side of the problem, the other side is how to use those signals for optimal trading.

Another stream of literature employs deep reinforcement learning to directly seek optimal trading policies, or strategies, that maximise profits. For example, \cite{ning2021double} use double deep Q-networks (DDQN) for optimal execution problems and \cite{casgrain2022deep} extends it to the multi-agent RL paradigm\footnote{The first versions of these works appeared online in 2018 and 2019.}; \cite{zhang2019deep} compares different paradigms of RL against classical time series momentum strategies on optimal futures trading, showing how RL-based methods outperform such baseline models; \cite{yang2020deep} trains an RL agent that makes use of an `ensemble' strategy --  an ensemble of trading strategies that uses three actor-critic based algorithms: Proximal Policy Optimisation (PPO), Advantage Actor Critic (A2C), and Deep Deterministic
Policy Gradient (DDPG). This  particular combined architecture enables the agent to inherit and integrate the best features of all three algorithms, thus robustly
adjusting to different market situations. \cite{cartea2023reinforcement} use DDQN and develop a new reinforced deep Markov model (RDMM) for statistical arbitrage.  Finally,  \cite{briola2023deep} start from an environment based on LOBs for a particular stock, and train a PPO agent to trade one unit of asset per time step, leading the agent to learn trading strategies that deliver stable positive returns in a highly stochastic and non-stationary environment.
This stream of literature is quite active  and thus many other methods have been developed to tackle other kind of problems when it comes to asset trading. For a more complete review of the methods available please refer to \cite{millea2021trading,zou2023stock, hambly2023recent}.

{To the best of our knowledge, the existing literature has paid limited attention to explicitly extracting latent state information from an observed trading signal and incorporating it into the training of an RL agent for optimal trading.
Thus, in this paper, we aim at filling this gap in the literature} by training an RL agent whose goal is to optimally trade a signal, hence maximizing their future discounted rewards from trading by making use of hidden information coming from the signal itself. 
We model the trading signal as a mean-reverting process; in this context, the information embedded in the training of the agent comes in the form of the probability of mean reversion to one of the different long-run regimes to which the signal mean reverts to. We progressively consider more complex signal dynamics and show how embedding information in the form of the probability of mean reversion to a regime improves the performance of the RL agent both in a simulated environment and with real data.
In this paper, we develop and test three DDPG-based algorithms to solve an optimal trading problem with mean-reverting signals governed by Markov chains. We incorporate GRU networks to capture the time-dependent structure of the trading signal, leveraging on this we explore different ways of feeding information to the RL agent. Our findings show that providing posterior probability estimates of the underlying mean-reversion regimes consistently yields the highest rewards, both in synthetic environments with multiple regime dynamics and in real market pairs-trading data among the three learning-based approaches. In contrast, supplying next-step signal forecasts adds little benefit, while using GRU hidden states provides intermediate performance. Overall, the results highlight that the quality and type of information provided to the agent is crucial: Interpretable structured insights into the data-generating process substantially improve the effectiveness and robustness of learning.
{{Rather than proposing a universal model-free trading framework, we focus on a class of trading problems in which profitability is driven by persistent but latent mean-reverting dynamics. In this setting, our objective is to investigate not only how a trading policy can be learned, but also how latent information should be represented before being supplied to the policy. In such problems, the relevant question is not only how to learn a trading policy, but how latent information should be represented before being passed to the policy and what is the gain in doing so, in terms of revenues from trading.}

The paper is organised as follows, in Section 2 we define the trading problem tackled throughout the remainder of the paper, in Section 3 introduces the methods used to model the trading problem, Section 4 discusses the results for different model scenarios with simulated data, Section 5 discusses the results obtained by applying the algorithms to real data, and finally, Section 6 provides conclusions and outlines further research directions.

\section{Optimal trading problem}
\label{sec:model}
{In this section we introduce the main trading problem, the market dynamics on which a trader is assumed to trade, and the partially observable specification of the resulting inter-temporal decision problem.}

\subsection{Market model}
\label{sec:market-model}

{We consider a risk-neutral investor who trades a mean-reverting signal $(S_t)_{t\geq0}$ with the objective of maximising expected discounted profits. The signal follows an  Ornstein-Uhlenbeck process with potentially time varying (and stochastic) parameters, thus it satisfies the stochastic differential equation (SDE):
\begin{equation}
\label{eqn:OU-SDE}
   \dS_t = \kappa_t (\theta_t- S_t) \,\dt + \sigma_t \, \dW_t \,,
\end{equation}%
where $W=(W_t)_{t\geq0}$ is a Brownian motion on a suitable probability space, and the parameters $\kappa_t\geq0$, $\theta_t>0$, and $\sigma_t>0$ are, respectively, the speed of mean reversion, the mean-reversion level, and the volatility of the signal at time $t$.

This class of trading problems arise often naturally in several trading applications, including statistical arbitrages, pairs trading and basis trading. In fact, in this class of trading problems the traders exploit the predictability coming from the tendency of the signal to revert towards its long run level. For instance, in the simple case where the parameters of Eq.\eqref{eqn:OU-SDE} are known, the sign and magnitude of the deviation $S_t - \theta_t$ provide information about future reversion of the signal, and it is exploited by the trader either by acquiring short or long positions on the signal itself, with the objective to maximise the revenues coming from this trading activity.  

The problem becomes substantially more challenging when the parameters are latent and time-varying. In this case,  \textit{hidden information} about the parameters plays a significant role. To illustrate this, consider first the case in which $\kappa_t=\kappa$ and $\sigma_t=\sigma$ are constant, while the mean-reversion level $\theta_t$ evolves according to some latent dynamics. Over an interval of length $\Delta t$, the solution of \eqref{eqn:OU-SDE} can be written as
\begin{equation}
\label{eqn:sde-sol}
    S_{t+\Delta t}
    =
    e^{-\kappa\Delta t}S_t
    +
    \kappa
    \int_t^{t+\Delta t}
    e^{-\kappa(t+\Delta t-u)}
    \theta_u\,\du
    +
    \sigma
    \int_t^{t+\Delta t}
    e^{-\kappa(t+\Delta t-u)}
    \dW_u .
\end{equation}

Denoting by $\F^S=(\F^S_t)_{t\ge0} :=  \left(\sigma\left(\{S_u\}_{u\leq t}\right)\right)_{t\ge0}$ the natural filtration generated by the observed signal, and taking conditional expectations in Eq.\eqref{eqn:sde-sol} we have
\begin{align}
\label{eqn:OU-relation}
    \E\left[
        S_{t+\Delta t}-S_t
        \mid \F_t^S
    \right]
    &=
    \left(e^{-\kappa\Delta t}-1\right)S_t
    +
    \kappa
    \int_t^{t+\Delta t}
    e^{-\kappa(t+\Delta t-u)}
    \E\left[\theta_u\mid\F_t^S\right]
    \du
    \notag\\
    &=
    \kappa\Delta t
    \left(
        \E[\theta_t\mid\F_t^S]-S_t
    \right)
    +
    o(\Delta t).
\end{align}
Hence, the expected signal increment depends on the trader's estimate of the latent mean-reversion level conditional on only the history of $S$ itself. When $\theta_t$ is not directly observed, the current value of $S_t$ alone is, in general, not sufficient to characterise the decision problem: the conditional expected increment of the signal depends on the conditional expectation of the latent mean-reversion level. This observation motivates the formulation of the trading problem as a partially observable decision problem. 
}

\subsection{Partial observability, decision problems, and belief states}

{To see how this problem is better formulated as a partially observable decision problem, we first consider the general decision making setting. In a fully observable Markov decision problem (MDP), the state $s_t \in \mathcal{S}$,  where $\mathcal{S}$ denotes the state space, contains all the information required to determine the conditional distribution of future states and rewards. Given a policy $\pi \in \Pi$, where $\Pi$ is the set of all feasible policies that map states to actions (we denote the set of allowed actions by $\mcA$), the value function is defined as  follows
\begin{equation}
\label{eqn:mdp-value}
    V^\pi(s)
    :=
    \E^\pi
    \left[
        \sum_{t=0}^{\infty}
        \gamma^t
        R(s_t,\pi(s_t))
        \,\middle|\,
        s_0=s
    \right],
\end{equation}
where $\gamma\in(0,1)$ is the discount factor and $R:\mathcal{S}\times\mcA\rightarrow\mathbb{R}$ denotes the one-step reward function, which assigns a scalar reward (or payoff) to each state-action pair $(s_t, \pi(s_t))$ given the policy $\pi$. In an MDP, the Markov property enforces that the future value depends on the history only through the current state, which ultimately leads to the Bellman equation
\begin{equation}
\label{eqn:mdp-bellman}
    V^\pi(s)
    =
    R(s,\pi(s))
    +
    \gamma\,
    \E^\pi
    \left[
        V^\pi(s_{t+1})
        \mid
        s_t=s
    \right].
\end{equation}

In a partially observable Markov decision process (POMDP), however, there is an additional underlying state $x_t\in\mathcal{X}$, where  $\mathcal{X}$ denotes the hidden or latent state space, which is not directly observed.
Instead, the agent receives observations generated from the hidden state and must use the available observation history to make inference about the hidden state. Consequently, the current observation alone is generally not sufficient to determine the conditional distribution of future states and rewards, as outlined in~\cite{aastrom1965optimal,kaelbling1998planning}.\footnote{A continuous time version of the problem is analysed by \cite{wonham1964some}.} The optimal decision problem then separates into filtering and optimisation, where the filtering depends on the history of the observable state $s_t$, i.e. $(s_u)_{u \le t}$. In this general formulation, we assume that the observable state coincides with the observed signal $S_t$.
In light of the dependency on hidden information, the problem of controlling a partially observable environment through the observable state $s_t$ leads to a modification of the Bellman equation. To this end, first denote
\begin{equation}
\label{eqn:general-belief}
    b_t(x) := \P\left(x_t = x \;|\; \F^S_t\right),
\end{equation}
which encodes the notion of \textit{belief} that the hidden state $x_t$ takes the value $x$ at time $t$ conditional on the filtration generated by the signal's history $(S_u)_{u\le t}$. The belief state is a sufficient statistic for the underlying hidden states and therefore allows the POMDP to be reformulated as an MDP whose state is extended by the belief distribution, as discussed in~\cite{smallwood1973optimal}, i.e., the POMDP becomes an MDP on the extended state space $\mathcal{S}\times \mathcal{P}(\mathcal{X})$, where $\mathcal{P}(\mathcal{X})$ denotes the set of probability measures over $\mathcal{X}$.

In this extended formulation, the value function in Eq.~\eqref{eqn:mdp-value} must account for the belief estimation, as the reward for an action $\pi(s_t)$ in state $s_t$ becomes the posterior expectation of the reward over the hidden states $x \in \mathcal{X}$, i.e., the expected reward is 
\begin{equation}
\label{eqn:belief-reward}
    \overline{R}(b,\pi(s_t))
    =
    \sum_{x\in\mathcal{X}}
    b(x)R(x,\pi(s_t)).
\end{equation}

Once the belief is included in the state, the conditional law of future observations and rewards is determined by the belief \textit{and} the observable variables. To this end, denoting by $b_t$ the belief that the hidden state takes value $x_t$, we have the modified Bellman equation
\begin{equation}
\label{eqn:pomdp-bellman}
    V^\pi(b)
    =
    \overline{R}(b,\pi(b))
    +
    \gamma\,
    \E^\pi
    \left[
        V^\pi(b_{t+1})
        \mid
        b_t=b
    \right].
\end{equation}

This belief-state formulation is the classical reduction of a POMDP to an MDP on the space of posterior distributions.
The distinction between Eq.~\eqref{eqn:mdp-bellman} and Eq.~\eqref{eqn:pomdp-bellman} is central to the problem considered in this paper. Under full observability, the current state is sufficient and the value function simply depends on it. Under partial observability, the relevant state for decision making also includes the agent's belief about the hidden state. Thus, the presence of latent information changes not only the policy but also the state space in which the value function is defined, making necessary a first estimation of the belief and, only after, a solution to the inter-temporal decision problem.
}

\subsection{The filtering and trading problem and its latent regime specification}

\paragraph{The MDP-POMDP specification.}
{We now specialise the general POMDP formulation, both for filtering and control, to the trading problem introduced in Section \ref{sec:market-model}. We assume that the mean reversion level $\theta_t$ is latent and evolves according to a Markov chain with $K$ states, while $\kappa$ and $\sigma$ are held constant. More specifically,
$$
\theta_t \in \{\phi_1, \dots, \phi_K\}.
$$
Let $A$ denote the generator matrix of the continuous time Markov chain governing $\theta_t$, i.e., the transition probability over a time interval of length $\Delta t$ is
\begin{equation}
\label{eqn:theta-transition}
    \P
    \left(
        \theta_{t}=\phi_j
        \mid
        \theta_{t-\Delta t}=\phi_i
    \right)
    =
    \left[e^{A\Delta t}\right]_{ij}.
\end{equation}

In the setting of Section \ref{sec:market-model}, the signal $S_t$ is observed and evolves according to Eq.~\eqref{eqn:OU-SDE}, while the mean-reversion level $\theta_t$ is latent and unobserved. Accordingly, the relevant belief state is the posterior probability assigned to each possible mean reversion-regime, and we denote the belief as:
\begin{equation}
\label{eqn:belief}
    \Phi_{t,k}
    :=
    \P
    \left(
        \theta_t=\phi_k
        \mid
        \F_t^S
    \right),
    \qquad
    k\in\K:=\{1,\ldots,K\},
\end{equation}
and respectively, the belief vector by $\Phi_t = (\Phi_{t,1},\ldots,\Phi_{t,K})$ -- a specialisation of the general belief state in \eqref{eqn:general-belief} to the finite-state latent regime $\theta_t$. According to Eq.\eqref{eqn:belief} the expected value of $\theta_t$ in Eq.~\eqref{eqn:OU-relation} may be written as
\begin{equation}
\label{eqn:theta-filtered-mean}
    \E[\,\theta_t\mid\F_t^S\,]
    =
    \sum_{k\in\K}
    \phi_k\,\Phi_{t,k}.
\end{equation}
Following the general framework for control with incomplete state information introduced by \cite{aastrom1965optimal}, the belief vector $\Phi_t$ may be updated recursively using a Bayesian filtering procedure; .

Letting $P = e^{A\Delta t}$ denote the transition matrix of the latent Markov chain, the prediction for the belief at time $t$ when observing states up to time $t-\Delta t$ is
\begin{equation}
\label{eqn:wPhi}
    \wPhi_{t,k} = \P(\theta_t =\phi_k\big| \F^S_{t-\Delta t}) = \sum_{j \in\K} \Phi_{t-\Delta t, j}\,P_{j,k}.
\end{equation}
This prediction step propagates the belief update from the previous time step according to the transition dynamics of the latent regime, without yet incorporating the information contained in the new observation $S_t$. Denoting by $\ell_k(S_t| S_{t- \Delta t})$ the likelihood of a new observation of the signal $S_t$ under hidden regime $k \in \K$, the \cite{aastrom1965optimal} filter gives the posterior update to the belief as: 
\begin{equation}
\label{eqn:belief-update}
    \Phi_{t,k}
    =
    \frac{
        \ell_k(S_t\mid S_{t-\Delta t})\,
        \wPhi_{t,k}
    }{
        \displaystyle
        \sum_{m\in\K}
        \ell_m(S_t\mid S_{t-\Delta t})\,
        \wPhi_{t,m}
    }.
\end{equation}

To efficiently use the output of the update rule of Eq.~\eqref{eqn:belief-update} in Eq.~\eqref{eqn:wPhi} to obtain a prediction for the belief, the transition rate matrix for the hidden states, along with the set of its possible values, and the likelihood of the observable process must be known to the agent. 

Having such hyper-parameter information is often unrealistic in real-world scenarios. Later, we use deep learning approaches to approximate this filter in settings where the hyper-parameters are not known.

\paragraph{The trading problem.}
Next, we discuss the trading problem. Here, the trader is allowed to trade at each time-step indexed as $t = 0, 1, 2 \dots$, each of length $\Delta t$, so that in real time the trader trades at $0,\Delta t, 2\Delta t, \dots$. At the beginning of the trading window she holds and initial inventory  $I_0\in\mathcal{I}:=\{\underline{I},\underline{I}+1,\dots,\overline{I}\}$. At each time-step $t$, her action is the amount of new inventory she holds and this persists until the next time step, thus $a_t = I_{t+1} \in \mcA=\{\underline{I},\underline{I}+1,\dots,\overline{I}\}$. Thus, at time $t$, the corresponding trading volume is $q_t = I_{t+1} - I_{t}$. Positive values of $q_t$ are purchases while negative values correspond to sales. 

We assume that when changing inventory incurs a proportional transaction cost $\lambda \ge 0$ per unit traded. Therefore  the one period reward has two components, the first from gains/losses due to changes in asset price and the second from transaction costs:
\begin{equation}
    \label{eqn:reward}
    R_t = I_{t+1}(S_{t+1} - S_{t}) - \lambda\big| q_t \big|.
\end{equation}
The reward depends on $S_t$, $S_{t+1}$, and $q_t$, while the conditional distribution of the future signal depends on the latent regime and, therefore, on the belief vector $\Phi_t$. 
The trader aims to maximise the expected discounted sum of all future rewards:
\begin{equation}
\label{eqn:max_pb}
    \max_{(q_t)_{t\geq0}\in\mathfrak{A}}
    \E
    \left[
        \sum_{t=0}^{\infty}
        \gamma^t R_t
        \,\Big|\,
        I_0,S_0,\Phi_0
    \right],
    \tag{P1}
\end{equation}
where $\gamma\in(0,1)$ is the discount factor and $\mathfrak{A}$ is the set of all $\F$-adapted policies of the form $q_t=q(I_t, S_t,\Phi_t)$ for some function $q:\mcI\times\R\times \mathcal{P}(\K)\to\mcI$, i.e., the set of policies consisting of maps from the current inventory, the current signal, and the belief state (a probability measure over the set of hidden states) to an action/inventory. 

The remainder of the paper studies  Problem~\eqref{eqn:max_pb} from a model agnostic viewpoint, i.e., without knowledge of the underlying Markov chain dynamics or values. We investigate how, and what kind of, information is effective in solving the POMDP. Moreover, we investigate whether information about the latent regimes may be inferred directly from the history of the observed signal without explicitly specifying the transition dynamics of the hidden process or the corresponding Bayesian filter. We do so in increasingly complex environment where the parameters $\kappa_t$ and $\sigma_t$ are modelled as independent Markov chains, therefore considering noisier environments where the correct estimate on mean-reversion levels, i.e. the hidden state, is crucial. We then use the resulting learned representation of the observable history and of the hidden states as an input to a deep reinforcement learning algorithm which learns the trading policy directly from the observed market dynamics. This reflects real scenarios where the data generating process (DPG) is completely or partially unknown. 

The remainder of the paper therefore focuses on the joint problem of \textit{learning} a suitable representation of the latent state and \textit{exploiting} such representation for trading under partial observability of the true state of the environment. To this end we propose three different algorithms that differ in how this information is obtained and then passed to the deep RL algorithm.  All with the goal of analysing how much and to what extent information passed to the agent matters in Problems like ~\eqref{eqn:max_pb}.
}

\section{Learning Algorithms}
\label{sec:algorithms}
{{In this section we present the deep RL algorithms that  we use to approximately solve the optimal trading problem. To simplify the presentation, in the following we detail the case where only the mean reversion level $\theta_t$ follows a Markov chain, while $\kappa$ and $\sigma$ are constant. The generalization to dynamic $\kappa$ and $\sigma$ is straightforward and the numerical results are shown for all models. }

As in any optimisation setting with latent and time-varying parameters, the problem faced by the algorithm is two-fold. The first consists in using the observed data to filter the latent parameters, while the second consists in finding the optimal action given the current estimate of the parameters. Here we consider and compare two approaches - one where we optimise the criterion without directly filtering the states of the system, and the second where we first develop a filter and then use the filtered states as part of the optimisation. The optimal trading problem can, therefore, be cast either as a unique optimisation or as two consecutive optimisations

Specifically, we first propose a one-step algorithm, using as features the path of the signal process and the current inventory and returns the optimal action - which treats the filtering and optimal trading problems simultaneously. We then consider a two-step approach, where we first train a model to determine posterior probabilities for the latent $\theta_t$ parameter from the paths of the observable signal process $S_t$, and then use the posterior probabilities, together with the current signal and inventory, to find the optimal trading action. As a third benchmark case, we also consider the case where the filtering part forecasts the next value of the trading signal, instead of the posterior probabilities, which is then used as feature in the part of the algorithm responsible of the trading action. In these last two settings the filtering and optimal trading are performed separately. Moreover, the setting where posterior probabilities are initially learned requires the knowledge of the true state of the latent process, a characteristic that is typically not available in real settings. 

For the architecture, we employ a Gated Recurrent Unit (GRU) network (introduced in \cite{cho2014learning}). The GRU network is a Recurrent Neural Network able to encode time dependency and thus deal with time series. One could also employ, e.g., long-short term memory processes or self-attention and other variants. {{We choose the GRUs over LSTMs as they have fewer gates, and hence fewer parameters, which lowers memory consumption and speeds up training within a larger architecture. Several empirical studies have found GRU architectures to outperform LSTM architectures in financial time-series forecasting, including cryptocurrency and stock-price prediction tasks, often while using fewer parameters and lower computational complexity, see, e.g., \cite{rekha2022sa}, \cite{kaur2025development}, and \cite{ghadimpour2026optimal}. A review on the GRU architecture in provided in Appendix\ref{appendix:GRU}. We analyse and justify this choice in a paragraph in Section \ref{sec:why-rnn}.}

To approximate the optimal trading policy we employ, as anticipated, Deep Reinforcement Learning, more specifically we use a Deep Deterministic Policy Gradient (DDPG) algorithm, first introduced by \cite{lillicrap2015continuous}, which employs an Actor-Critic approach. The algorithm makes use of two distinct neural networks, an Actor network $\pi$ that is responsible to choose the action to perform, based on the state of the environment, and a Critic network $Q$ that evaluates the `quality' of the action chosen using the network $\pi$ given the state of the environment. 
For each of the algorithm proposed we train the RL agent over a number $N$ of training episodes.
In our setting, the agent is tasked with the objective of maximising the rewards obtained by rebalancing the inventory holdings $I$ as 
defined in Eq. \eqref{eqn:reward}. The agent's actions consist of the new level of inventory to be held. Thus, the agent seeks to learn the optimal rebalancing their level of inventory depending on the signal process $S_t$. After the $N$ trading episodes we test what the agent has learnt by feeding other $M$
episodes, where the agent trades using the policy learnt during the training phase.

\subsection{States, environment, actions, and rewards}
\label{sec:actions_rewards}
While training, the agent may be at some time \(t\) with an inventory \(I_t\) and must decide how much inventory to hold at time \(t+1\), i.e., \(I_{t+1}\). The agent has access to the signal value at that time \(S_t\) and to the \(t-W\) past observations of the signal, where $W+1$ is the length of the signal window, which we denote $\{S_{u}\}^{t-W}_{u=t}$.
A visual representation of the information  available to the agent is shown in Fig.~\ref{fig:signal_path}. These features/information are used to train an agent to optimise their sum of discounted rewards  using a GRU network and two variants of the DDPG algorithm - i.e., we consider two approaches to the joint problem of filtering the signal and finding the optimal trading strategy employing this set of information about the environment available to the agent.

For training purposes, we simulate batches of size $b$ of time series for the signal $\{S_{u}\}^{t+W}_{u=t}$ (i.e., signal time-series of length $W+2$) and inventories $I_t$. Inventories $I_t$ are randomly sampled as $I_t  \sim {\mathcal U}_{[I_{\text{min}}, I_{\text{max}}]}$.\footnote{Where $\mathcal U$ is the uniform distribution.}

For the signal, we simulate the time series of the signal \(S_t\) of length \(W+2\), with parameters according to the setting being considered. The starting value for the signal's time series simulation is $S_{t-W} \sim  \mathcal{N}(\mu_{\text{inv}},3\,\sigma_{\text{inv}})$ 
where $\sigma_{\text{inv}} = \frac{\sigma}{2 \kappa}$ which is the invariant volatility for the Ornstein-Uhlenbeck process with {{constant} parameters, while $\mu_{\text{inv}}$ is the invariant mean of the {{constant parameter version of the trading signal in Eq.~\eqref{eqn:OU-SDE}}.\footnote{In the settings where \(\sigma\) and \(\kappa\) are time varying in order to compute the initial value of \(S_0\), we choose the minimal value according to the regimes used.}

\begin{figure}[h]
\centering
\begin{tikzpicture}[scale=1.5]

\draw[thick,->] (0,0) -- (5,0) node[anchor=west] {\small Time};
\draw[thick,->] (0,0) -- (0,2.2) node[anchor=south, rotate=90, xshift=-40pt] {\small Signal Value};

\foreach \x in {0,4,4.1}
  \draw[red,ultra thick] (\x,3pt) -- (\x,-3pt);

\node at (0,-0.25) {\small $t - W$};
\node at (3.95,-0.25) {\small \textcolor{darkgreen}{$t$}};
\node at (4.25,-0.25) {\small $t + 1$};

\draw[dashed,gray] (0,0) -- (0,2);
\draw[dashed,gray] (4,0) -- (4,2);
\draw[dashed,gray] (4.1,0) -- (4.1,2);

\pgfmathsetseed{1123}
\draw[thick,blue!70!black,opacity=0.9] (0,0.5)
  \foreach \x in {1,...,40}
    { -- ++(0.125,rand*0.2) };

\filldraw[blue] (4,1.252) circle (1.2pt) node[anchor=north east] {\small $S_t$};
\filldraw[red] (4.1,1.2) circle (1.2pt) node[anchor=south west] {\small $S_{t+1}$};

\draw [decorate,decoration={brace,amplitude=6pt},thick,darkgreen]
  (0.05,2) -- (4,2) node[midway,above=6pt] {\small $\{S_u\}_{u = t}^{t-W}$};

\end{tikzpicture}
\caption{{A schematic of the signal information that the agent may use.  At time \(t\), the agent} has access to the past values of the signal from time \(t\) back to time \(t-W\) and decides how much inventory to be held at time \(t+1\).}
\label{fig:signal_path}

\end{figure}

In the testing phase, where we assess the trading policy learnt by the agent, we assume that the agent trades over a time-window of $n$ time-steps. 
We assume that the agent starts their trading at time $t =  0$ and has access to information on the past $W$ values of the signal $S$.  As time $t$ progresses, the agent uses the observations of the signal from time $t$ back to time $t-W$, in a rolling window fashion, to decide the inventory to be held at time $t+1$.
The agent can hold inventories within the interval $I\in[-I_\text{max}, I_{\text{max}}]$ and we always start the testing episodes of trading with $S_0 = 1$ and $I_0=0$.

As mentioned, we employ two main approaches: In the \emph{one-step approach} we set the states, or information about the environment, available to the agent as tuples of
$(S_t, I_{t}, o_t)$.
Here $S_t$ is the signal value at time $t$, $I_t$ is the value of the inventory held at time $t$, and $o_t$ is the output of a Recurrent Neural Network that takes as input a collection $\{S_{u}\}^{t-W}_{u=t}$
of past values of the signal from time $t-W$ up to time $t$ (see below for more details).

In the \emph{two-step approach} we decouple the filtering from the optimal trading task and consider two different settings. In the first setting, we first train the algorithm to learn the regimes from the past and current value of the signal. Specifically, the posterior probabilities $ \Phi_{t,k} := \P(\theta_t=\phi_k| \{S_{u}\}^{t-W}_{u=t})$ and $k = 1,2,3$ are learnt offline using a GRU network followed by a feed-forward network with soft-max activation output layer (to provide estimates of the posterior probability of the regime based on the historical price signal). Then the DDPG uses the tuple $(S_t, I_t, \{\Phi_{t, k}\}_{k=1,2,3})$ as features to learn the optimal trading. In the third setting, we first train the algorithm to forecast the next value of the signal $\tilde S_{t+1}$ and then the DDPG takes as input features the tuple $(S_t, I_t, \tilde S_{t+1})$.

The action that the agent can take at each time step $t$ consists of a rebalancing of her inventory, which can be a long ($I_{t+1}>0$) or short ($I_{t+1}<0$) position, and is chosen according to the algorithms described below. As a constraint, we impose that at each time, the agent holds $I_t \in  [I_{\text{min}}, I_{\text{max}}]$,  i.e. the actions that can be chosen limit $I_t$ between a maximum and a minimum inventory. 

After the action has been taken by the agent in a state of the environment at time $t$, the reward is calculated considering the subsequent change in the signal process and measuring the gain obtained by the agent trading.
The reward used is given by Eq. \eqref{eqn:reward}.

In all algorithms, we let the agent explore a wide range of combinations of inventory holdings $I_t$ and possible $S_t$ values or, in the case of the two-step procedures, multiple combinations of states $(I_t, S_t, \Phi_{t, k})$ or $(I_t, S_t, \tilde S_{t+1})$. This is done through an exploration-exploitation scheme. 
Thus, during the learning procedure, the agent employs randomized actions, which is a randomized perturbation of the current policy, the exploration process controlled by an exploration parameter \(\varepsilon\) that decays as the training unfolds.

\subsection{The one-step approach}
\label{sec:methodS_one_step}

To approximate the optimal strategy for Problem~\eqref{eqn:max_pb} when the dynamics of the trading signal are described by Eq.~\eqref{eqn:OU-SDE} with regime switching parameters, we first propose a one-step procedure that employs the signal's past history as part of the state of the environment that feeds into a DDPG algorithm, employing the Actor-Critic approach. To account for the time dependency of the trading signal, we encode this information using a GRU network whose output is then fed into the DDPG algorithm. A directed graph representation of the one-step approach is reported in Fig.~\ref{fig:one-step-architecture}. 

\input{onestep}

\subsubsection{The training loop}

The algorithm is trained over $N$ iterations, for each iteration we sample $b$ batches of input data that are used to train the algorithm. 
In each iteration, both the Agent and the Critic network are updated by feeding to the networks independent batches of states of the world represented by a matrix $\f \in {\mathbb R}^{b \times (W + 2)}$, where each of the $b$ rows contains a tuple of the form $(\{S_{u}\}^{t-W}_{u=t}, I_{t})$. 

\paragraph{The Actor and Critic networks.}
The Critic and Actor neural networks are feed-forward nets with four and three input neurons, respectively, a number of $l_{\text{NN}}$ layers of $d_{\text{NN}}$ hidden nodes, each with \verb!SiLu! activation function, in the Actor network, the final layer has \verb!tanh! activation function, whose output is then scaled by $I_{\text{max}}$. 

As we are not assuming any initial or final inventory goals or penalties, in our DDPG implementation, while training, we choose not to keep memory of the states that the agent has visited during training. Rather, we feed randomly generated initial states of the world in order to let the algorithm explore as many states as possible, as discussed in Sec.~\ref{sec:actions_rewards}.  

This is done to provide the agent with as many different states as possible when training. Moreover, since we do not consider any form of permanent impact generated by the agent when trading, the buy and sell actions performed by the agent have no direct effect on the signal process, and therefore no memory needs to be kept.

At the beginning of the training loop we initialise the networks $\pi, Q$ with random weights $\mu_\pi, \mu_Q$. As training unfolds, $(\mu_\pi, \mu_Q) \to (\mu^*_\pi, \mu^*_Q)$, where $(\mu^*_\pi, \mu^*_Q)$ are the optimal weights that correspond to the optimal policy that maximises the sum of discounted rewards. 
 
During training the $Q$ network, in order to avoid an overestimation of the $Q$-values for the considered state-action pairs, we use a $Q_{\text{tgt}}$ network that is initialised at the beginning of the training routine as a copy of the $Q$ network with weights $\mu_{Q_{\text{tgt}}} = \mu_{Q}$, thus employing double deep Q-learning for the Critic part of the algorithm, the weights of the $\mu_{Q_{\text{tgt}}}$ are then periodically set equal to $\mu_Q$. In order to do so, we have adopted the soft update technique as described in the original DDPG paper by \cite{lillicrap2015continuous}, with the soft update parameter set to $0.001$.

\paragraph{The Critic network.}

For each training iteration, we train a Critic network $Q$. To this end, we add an additive exploration zero mean noise to the output of the Actor network $\pi$ (whose training is described below), whose variance declines at each iteration\footnote{Specifically, at the learning loop start we set $a > 0$ and $\varepsilon_{\text{min}} < a$ 
Then, for each train iteration $m \in \{1,N\}$ we set $\varepsilon = \max(a/(a + m), \varepsilon_{\text{min}})$. In this way, the more we iterate in the learning routine the lower the $\varepsilon$ value. $\varepsilon$ is then used as an additive noise in the Actor-Critic architecture used in both the learning algorithms.}.
The additive noise, $\varepsilon$, helps the Actor neural network to explore the range of inventory holdings to be held depending on the state.
Thus, while training the $Q$ network, $I_{{t+1}} = \pi(\g_t|\mu_\pi) + \mathcal{N}(0, \varepsilon)$, where \(\pi(\g_t|\mu_\pi)\) is the output of the Actor network. Clearly, as the training unfolds and $\varepsilon$ becomes smaller, the exploration noise diminishes and the policy chosen by the network $\pi$ becomes deterministic. In this way, given the weights $\mu^*_\pi$, found at the end of the training, the Critic network directly maps a state into an action such that:
\begin{equation}
    I^{*}_{{t+1}} = \pi(\s_{t}|\mu^*_\pi)
\end{equation}
Once the new level of inventory is obtained, it is used to calculate the reward for each state in the batch as in Eq.~\eqref{eqn:reward}.

To train the $Q$ network, we update the weights $\mu_Q$ such that $\mu_Q = \argmin_{\mu_Q} \L_1(\mu_Q; \mu_{Q_{\text{tgt}}})$, by taking a single gradient step on the loss 

\begin{align}
        \L_1 &= \frac{1}{b} \sum^b_{i=1} (Q(\g_t^{(i)}, I^{(i)}_{{t+1}}|\mu_Q) - y^{(i)}_t)^2\\
        y^{(i)}_t &= r^{(i)}_{t} + \gamma Q_{\text{tgt}}(\g'\,^{(i)}_{t+1}, \pi(\g'\,^{(i)}_{t+1}|\mu_\pi)|\mu_{Q_{\text{tgt}}})
\end{align}

where the superscript $i$ indicates the $i$-th row of the matrix $\g_t$ and $\g'_t \in \R^{b\times (W+2)}$ is the matrix of inputs \(\g'_{t+1} = (o_{t+1}, S_{t+1}, I_{t+1})\), where \(o_{t+1}\) is obtained by passing to the GRU network the series of $\{S_{u}\}^{t-W+1}_{u=t+1}$ 
of past values of the signal from time $t-W+1$ up to time $t+1$.
The weights of the Critic network $\mu_Q$ are then updated using gradient descent.

Within the training loop we iterate this training process for the Critic network a number $\ell$ of times in order to explore the space of actions given the states more thoroughly, in order to  obtain better estimates for the $Q$-values.

\paragraph{The Actor network.}
For each training iteration we train the Actor network $\pi$, which is the neural network responsible for choosing the action $I_{t+1}$. 
To do this, we feed the $\pi$ network with a batch of $b$ states $\g_t$.  The $\pi$ network returns a new level of inventory to be held $I_{t+1}$, which is in turn fed to the Critic $Q$ network, along with the state $\g_t$. The output of the $Q$ network is the $Q$ value which is maximized during training. Specifically, the weights $\mu_\pi$ of the Actor network are found by maximizing the output of the Critic network with fixed weights $\mu_Q$. As before we perform a single gradient step to minimise the loss:

\begin{equation}
    \L_2 = - \frac{1}{b} \sum^b_{i =1} Q(\g_t^{(i)}, \pi(\g_t^{(i)}|\mu_\pi)|\mu_Q)
\end{equation}
 where $\pi(\g_t^{(i)}|\mu_\pi)$ is the output of the Actor network with current weights $\mu_\pi$.
The weights $\mu_\pi$ are updated using the policy gradient theorem. The derivative of the loss function $\L_2$ with respect to the weights of the Actor network is:
\begin{equation}
    \nabla_{\mu_\pi}\L_2 = - \frac1b\sum^b_{i=1}\left[ \nabla_{a}Q(\s_t^{(i)}, a^{(i)}|\mu_Q)|_{a^{(i)} = \pi(\s_t^{(i)}|\mu_\pi)} \nabla_{\mu_\pi}(\s_t^{(i)}|\mu_\pi) \right]
\end{equation}
where the first part in the square brackets tells us how sensitive the $Q$ network is to changes in the action chosen by the Actor $\pi$, this is the gradient of the $Q$-value with respect to the action and evaluated at the action suggested by the Actor network, in this context for ease of notation we use $a^{(i)}=I^{(i)}_{{t}+1}$ to denote the action for the \(i\) state among the $b$ considered in the batch. The second part inside the brackets is the gradient of the actor's policy and accounts for how much the action $I_{t+1}$ changes with respect to changes in the $\pi$ network parameters. We update the weights for the $\pi$ network with gradient descent. We repeat this training routine a number $l$ of times in order to assess how the quality of the actions chosen by the Actor changes and, in this way, to thoroughly explore weights combinations for the $\pi$ network. 

The complete training algorithm for both the Critic and the Actor neural networks is reported in Algorithm~\ref{algo:ddpg}. Both the Actor and the Critic neural networks are feed-forward fully connected neural networks, for the gradient descent we use the Weighted ADAM optimiser with a scheduler. The features for the DDPG algorithm are normalised in the domain $[0, 1]$. Relevant parameters for the networks and the market model are reported in Table~\ref{tab:sim_pmts}. We will refer to this model as \textit{hid-DDPG} throughout the paper.

\subsection{The two-step approach}
\label{sec:methodS_two_step}

To approximate the optimiser of  Problem~\eqref{eqn:max_pb} when the trading signal follows the dynamics in Eq.~\eqref{eqn:OU-SDE} with regime switching parameters, we further propose a two-step procedure in two different setups. For the first setup, in the first step, we approximately solve the filtering problem, aiming at estimating the posterior probability of being in one of the regimes of $\theta_t$. In the second step, we approximate the optimal trading policies that use the estimated posterior distribution as an additional feature. 
These estimates, together with the signal $S_t$ and the inventory level $I_t$, constitute the features used in the RL algorithm. The directed graph representation of the training for this algorithm is reported in Fig.~\ref{fig:two-step-architecture}. In the third setup we repeat the procedure but, instead of filtering the posterior probabilities for the $\theta_t$ regimes, we train a GRU to obtain estimates of the next value of the trading signal, i.e.  $\tilde S_{t+1}$, we will use this estimate along with the signal $S_t$ and the inventory level $I_t$ as features to the DDPG algorithm. 
The directed graph representation of the training for this algorithm is reported in Fig.~\ref{fig:two-step-architecture2}. 
The DDPG step for both the two-step procedure remains the same, except for the set of features used, as discussed at the end of Sec.\ref{sec:actions_rewards}.

\subsubsection{First-step: regime filtering with classification}
\label{sec:regime_filter_method}

Concerning the first step, we approximately solve the filtering problem, aiming at estimating the posterior probability of being in one of the regimes for $\theta_t$. Further, we approximate the optimal trading policies using the estimated posterior distribution as an additional feature. 

\input{twostep}

In this setup, we provide a deep learning approach to estimate the probability that $\theta_t$ is in one of the regimes, conditional on the history of the path of the signal process $S$. That is, we aim to find an approximation
\begin{equation}
\label{eqn:approx_pbm}
    \Phi_{t,k} := \P(\theta_t=\phi_k| (S_u)_{u\le t}) \approx g_k\left( S_t, \dots , S_{t-W}\right)\,,\qquad k=1,2,3
\end{equation}
The vector valued function $g(\cdot)=(g_1(\cdot),g_2(\cdot),g_3(\cdot))$ is approximated as an Artificial Neural Network (ANN) with $W$ being the look-back window used. The input, $\{S_{u}\}^{t-W}_{u=t}$, to the ANN $g(\cdot)$ is a sequence of trading signals that goes from time-step $t$ backwards to time $t-W$.

Once again, as the signal $S_t$ is a time series, a natural choice is to employ RNNs in the approximation consistently with the one-step method outlined in Sec.~\ref{sec:methodS_one_step} - specifically, we employ a GRU network as our function approximant $g(\cdot)$.
On top of the GRU architecture, we nest additional fully connected layers  to return the probability for the signal $S_t$ to be mean reverting to a specific regime $\theta_t \in\{\phi_1,\phi_2, \phi_3\} $.

More specifically, the last hidden state is passed to a second architecture equipped with additional fully connected layers. In each hidden layer of the fully connected neural net we apply the Sigmoid Linear Unit (\verb!SiLu!) activation function to add non-linearity. In the end, since this is a classification problem, we use the \verb|SoftMax| activation function to retrieve the probability $\P\left(\theta_{t}=\phi_j|\{S_{u}\}^{t-W}_{u=t}\right)$. The loss used to train the architecture is the categorical cross entropy. 

The parameters for the GRU and the additional layers are reported in Table~\ref{tab:sim_pmts}.

\subsubsection{First-step: regression setup}
\label{sec:regression setup}

As a point of comparison we develop a third setup to the two-step procedure. In this third experiment, instead of approximating the probability of being mean reverting at each of the regimes of $\theta_t$,  the first step consists in training a GRU net on a regression problem to directly estimating the next value for the trading signal, i.e. $\Tilde{S}_{t+1}$. 
Similarly as in the filtering problem, in order to obtain the next value estimate, we provide as input to the GRU $g(\cdot)$ a sequence $\{S_{u}\}^{t-W}_{u=t}$.
The architecture of the GRU net remains the same but the last layer has \verb!SiLu! activation function and we minimise the loss consisting of the squared error between the estimate $\Tilde{S}_{t+1}$ and the actual next value of the trading signal $S_{t+1}$. 

\input{twostep_reg}

\subsubsection{Second-step: optimal trading}

Once training in the first-step is completed, and the GRU models are trained for either of the setups in \ref{sec:regime_filter_method} or \ref{sec:regression setup}, we consider the problem of a risk neutral trader who maximises the expected discounted profits from trading. We assume that the agent re-balances the inventory according to the level of the observed trading signal and on their current inventory in order to maximise the rewards obtained from trading. 
To do so, we again model the agent with a DDPG algorithm similarly as in Sec.~\ref{sec:methodS_one_step}. Hence, for the optimal trading part of the algorithm we again adopt an Actor-Critic approach where we make use of two distinct neural networks, an Actor network $\pi$ that is responsible for choosing the action to perform, based on the state of the environment where the agent is, and a Critic network $Q$ that critically evaluates the `quality' of the action chosen using the network $\pi$ given the state of the environment. 
This is analogous to the one-step procedure with the important difference that, in the regime filtering approach, the posterior probabilities estimates on the current mean reversion regime $\theta_t$ are added to the feature set and the model that approximately solves the filtering model is trained in the previous step, while in the regression approach the estimate for the next value of the trading signal $S_t$ is added to the set of features fed to the RL agent. 

Hence, the first setup (and second method proposed) of the two-step approach is termed {\it prob-DDPG} and makes use of features $\s_{t} = (S_{t}, I_{t}, \Phi_{t, k})$ for the DDPG part, while the second setup (and third method overall) is termed {\it reg-DDPG} throughout the paper, and employs $\s_{t} = (S_{t}, I_{t}, \Tilde{S}_{{t+1}})$ as features to the DDPG. 

For the DDPG part we again set both Actor and Critic ANNs to be feed-forward fully connected neural networks as in the one-step approach. For the gradient descent steps we use the Weighted ADAM optimiser with a scheduler. The features of the DDPG algorithm are normalised in the domain $[0, 1]$. The relevant parameters for the networks and the market model are reported in Table~\ref{tab:sim_pmts}.

\subsubsection{{Why a recurrent neural network}}
\label{sec:why-rnn}
{
The discussion in Section~\ref{sec:model} shows how having only access to the history of the observable signal, and not the latent state, modifies the decision problem.  When the dynamics (and not only the value) of the latent state is unknown, a question that arises is how to efficiently use the signal's history to construct a suitable latent state belief representation which may then be used to learn approximately optimal policies. This motivates the use of a recurrent neural network (RNN). An RNN provides a parametrised representation of an ordered sequence of data, in our case, $(S_u)_{u \le t}$, through an internal state that is recursively updated as new observations become available. Schematically, this hidden RNN state may be written as 
\begin{equation}
    h_t = f_{\vartheta}(h_{t-1}, S_t),
\end{equation}
where $h_t$ summarizes information extracted from the observations up to time $t$ and $\vartheta$ denotes the RNN's trainable parameters. The network does not treat each observation independently, but rather recursively updates its representation as new observations become available (see: \cite{chung2014empirical}).  This property makes recurrent architectures particularly suitable for sequential decision problems in which information relevant for the current decision is distributed over time (see: \cite{ lambrechts2022recurrent}).

In our setting, the connection between an RNN and the Bayesian filtering introduced in Section \ref{sec:model} is natural. The filter in Eq.~\eqref{eqn:belief-update}
may be rewritten  as $\Phi_t = F(\Phi_{t-1}, S_{t})$, where $F(\cdot)$ is a map determined by the transition and observation model, along with the past history of the signal $\F^S_t$. From Eq.\eqref{eqn:belief-update}, we see that the exact filter combines information contained in the previous belief distribution with the likelihood of a new observation. This recursive structure is replicated by the recursive structure of the RNN; however, the RNN is model-agnostic in that none of the parameters that generate the dynamics of the hidden state enter explicitly into the RNN parameters; rather, they are learned from the data itself. As such, the RNN acts as a model-free \textit{approximation} of the filtering map. 
The RNN's output layer is further mapped through a \texttt{softmax} activation function  resulting in values in a probability simplex. Thus, the output may be interpreted as an estimated distribution over the possible regimes.  

This interpretation is closely related to the use of RNNs as data-driven approximations of Bayesian filters (see: \cite{bishop2023rnn}). Thus, the \textit{prob-DDPG }algorithm outlined above may be seen as a model-agnostic approximation to belief filtering.

In forecasting-based trading strategies, GRU networks are widely used to predict the next signal value, $S_{t+1}$ (see, e.g., \cite{ghadimpour2026optimal}). More generally, predicting the next signal value and using this forecast to inform trading decisions is a common approach when the DGP is unknown, providing further motivation for the \textit{reg-DDPG} approach introduced above.
}
{
\paragraph{Learning the belief representation.} 
We now specialise the GRU to estimating a belief representation, and  justify why it approximates the posterior probability estimation for the latent regimes.
First, recall that the target of the GRU is, more specifically, the conditional distribution with respect to $W$ past observations of $S_t$ 
\begin{equation}
    \Phi_{t,k} = \P \left( \theta_t = \phi_k \big| (S_u)^{t-W}_{u = t}\right)
\end{equation}
and we denote the output of the GRU by
\begin{equation}
    \hat{\Phi}^{\vartheta}_t = g_{\vartheta}\left(
        S_{t-W},\ldots,S_t
    \right).
\end{equation}

The categorical cross entropy loss used in the first step of \textit{prob-DDPG} can be written as
\begin{align}
\label{eqn:cross-entropy}
    \LL(\vartheta) 
    &= - \E\left[\sum_t\sum_k \mathds{1}_{\left\{\theta_t = \phi_k\right\}}\log\hat{\Phi}^{\vartheta}_{t, k}\right]
    \\
    &= 
    - \E\left[\sum_t\sum_k \E\left[ \mathds{1}_{\left\{\theta_t = \phi_k\right\}}\log\hat{\Phi}^{\vartheta}_{t, k}\,\big|\,\F_t^S\right]\right]
    \\
    &= 
    - \E\left[\sum_t\sum_k \E\left[ 
    \mathds{1}_{\left\{\theta_t = \phi_k\right\}}\,\big|\,\F_t^S\right]\log\hat{\Phi}^{\vartheta}_{t, k}\right]    
    \\
    &=
    - \E\left[\sum_t\sum_k \Phi_{t,k}\,\log\hat{\Phi}^{\vartheta}_{t, k}\right]        
\end{align}
This can furthermore be decomposed into the sum of Kullback-Leibler (KL) divergence from $\hat{\Phi}^{\vartheta}$ to $\hat{\Phi}$ and the entropy of $\Phi$ as follows
\begin{align}
    \LL(\vartheta)
    &=
     \E\left[\sum_t\sum_k \Phi_{t,k}\,\left(\log\frac{\Phi_{t,k}}{\hat{\Phi}^{\vartheta}_{t, k}} - \log\Phi_{t,k} \right)\right]
    \\
    &=
    \mathbb{E}
    \left[\sum_t
        \mathrm{KL}
        \left(
            \Phi_t
            \,\middle\|\,
            \hat{\Phi}^{\vartheta}_t
        \right)
    \right]
    +
    \mathbb{E}
    \left[\sum_t
        H(\Phi_t)
    \right],
    \label{eqn:loss_decomposed}
\end{align}
where
\begin{equation}
    H(\Phi_t)
    =
    -\sum_k
    \Phi_{t,k}\log\Phi_{t,k}
\end{equation}
denotes the entropy of the true posterior distribution.

The second term in Eq.\eqref{eqn:loss_decomposed} does not depend on the network parameters; therefore, minimising the categorical cross-entropy in Eq.\eqref{eqn:cross-entropy} is equivalent to minimising the expected KL divergence between the estimated and the exact posterior distributions. Further, one can show that
$\E [\mathrm{KL}
    (
        \Phi_t
        \,\|\,
        \hat{\Phi}^{\vartheta}_t
    )]
=0$
if and only if the estimated posterior distribution \textit{coincides} with the exact one almost surely.  This shows that minimising the cross-entropy \eqref{eqn:cross-entropy} leads to $\hat{\Phi}^\vartheta$ closely approximating the true posterior $\Phi$.
}
{
\paragraph{The look back window $W$.} We now turn to the question of choosing the look-back window $W$. The exact belief  defined in Eq.~\eqref{eqn:general-belief} is an expectation conditional on the full history $(S_u)_{u\le t}$ of the observable process at time $t$. We, however, truncate the GRU lookback window to size $W$  for computational efficiency, and this introduces a further gap between $\hat{\Phi}^{\vartheta}_{t,k}$ 
and the exact filter. Under suitable filter-stability conditions, finite-memory policies are known to be near-optimal for POMDPs (see: \cite{kara2023q}). This motivates treating the window length as an important design parameter.

To assess the effect of different window lengths on the GRU, we consider $W\in\{5,10,20,50,100\}$
and compare the resulting GRU estimates, $\hat{\Phi}_{t,k}$,  with the posterior produced by the exact filter in Eq.~\eqref{eqn:belief-update}. We use two measures of discrepancy: the mean KL divergence and the total variation (TV) distance, defined path-wise as 
\begin{equation}
    \label{eqn:TV}
    \operatorname{TV}
    \left(
        \Phi_t,\hat{\Phi}_t
    \right)
    =
    \tfrac{1}{2}
    \sum_k
    \left|
        \Phi_{t,k} - \hat{\Phi}_{t,k}
    \right|.
\end{equation}
We estimate these quantities on 250 out-of-sample paths, each consisting of 300 time steps. The results are reported in Table~\ref{tab:GRU_W}.
\begin{table}[H]

    \centering
    \caption{{Agreement between the trained GRU posterior $\hat\Phi_t$ and the exact filter $\Phi_t$ Eq.\eqref{eqn:belief-update}, as a function
    of the input window length $W$, computed on out-of-sample paths in two settings where(i) only $\theta_t$ is a MC and (ii)  $\theta_t, \kappa_t, \sigma_t$ are MCs. Standard errors across paths in parentheses.}
}    \label{tab:GRU_W}
    \begin{tabular}{@{}c c c@{}}
        \toprule
        \multicolumn{3}{c}{$\theta_t$ are MCs}
        \\
        \toprule
        Window $W$ & TV distance & Mean KL \\
        \midrule
        $5$   & $0.155\ (0.010)$ 
        & $0.143$ \\
        $10$  & $0.091\ (0.007)$          & $0.068$ \\
        $20$  & ${0.070}\ (0.006)$ & ${0.061}$ \\
        $50$  & $0.097\ (0.009)$          & $0.222$ \\
        $100$ & $0.121\ (0.010)$          & $0.170$ \\
        \bottomrule
    \end{tabular}
    \qquad
    \begin{tabular}{@{}c c c@{}}
        \toprule
        \multicolumn{3}{c}{$\theta_t$, $\kappa_t$, and $\sigma_t$ are MCs}
        \\
        \toprule
        Window $W$ & TV distance & Mean KL \\
        \midrule
        $5$   & $0.152\ (0.009)$          & $0.135$ \\
        $10$  & $0.092\ (0.007)$          & $0.065$ \\
        $20$  & ${0.068}\ (0.006)$ & ${0.041}$ \\
        $50$  & $0.082\ (0.007)$          & $0.061$ \\
        $100$ & $0.091\ (0.012)$          & $0.082$ \\
        \bottomrule
    \end{tabular}
\end{table}

The results reported in Table~\ref{tab:GRU_W}  show that intermediate window lengths provide the best agreement with the exact filter.
For instance, when only $\theta_t$ switches, the windows $W=10$ and $W=20$ provide the lowest discrepancies with the exact filter. When $ \theta_t$, $\kappa_t$, and $\sigma_t$ are all allowed to switch, the best performance is again obtained for $W=20$. These results suggest that providing additional observations to the network does not monotonically improve its approximation capabilities with a fixed training budget. Moreover, very short windows seem  not to contain enough information to match the true posterior. 

Although $W=20$ provides the best approximation to the exact filter, we use $W=10$ in the \textit{prob-DDPG} and \textit{hid-DDPG} methods in the less complex environments, as it provides a more parsimonious representation while maintaining comparable downstream performance and preventing the risk of overfitting. When $\kappa_t$ and $\sigma_t$ also switch, three latent chains must be inferred from a single scalar observation, and the two additional parameters are identifiable only from the dynamics of the signal across several increments. Accordingly, the stochastic reversion speed and volatility make the problem of inferring the  reversion level from a short observation history generally more difficult. Therefore, in these settings we use $W = 20$ as suggested by the results in Table ~\ref{tab:GRU_W}.

Figure~\ref{fig:gru_aastrom} provides a representative comparison between the exact Bayesian filter and the GRU approximation in the simplest setting. Despite the fact that the GRU \textit{does not have access} to the transition and observation models used by the exact filter, its estimated posterior closely tracks the exact filtering probabilities. This provides evidence that the GRU is able to recover a substantial part of the filtering information directly from the observable signal.
\begin{figure}
    \centering
    \includegraphics[width=0.7\linewidth]{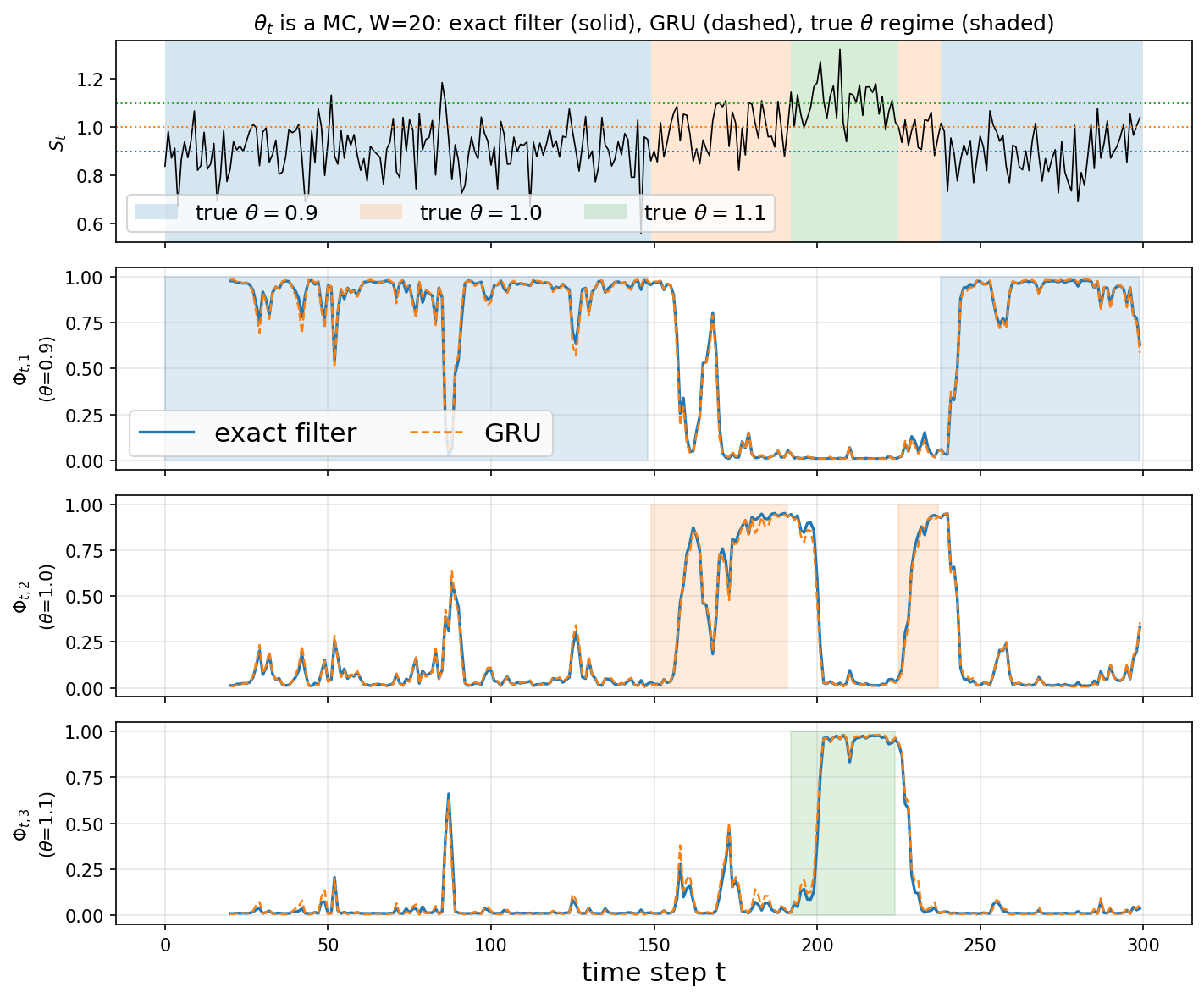}
    \caption{Estimated posterior probabilities for the $\theta_t$ regimes in the simplest case where $\theta_t \in \{ \phi_1, \phi_2, \phi_3\}$ is a Markov chain together with the filter in Eq.\eqref{eqn:belief-update}(solid blue) and the GRU approximator (dashed orange).}
    \label{fig:gru_aastrom}
\end{figure}

For the \textit{reg-DDPG} approach, the first-stage objective is to forecast the next value of the observed signal. While inferring the current regime requires discriminating among a finite number of possible latent configurations, predicting $S_{t+1}$ requires learning the conditional dynamics of the signal, primarily through the reversion level and speed, while volatility contributes to the uncertainty associated with the next observation. Moreover, the next observation contains a diffusive innovation that is unpredictable and uncorrelated with the regime changes. Our findings show that the forecasting task requires a much longer historical window of size  $W = 50$ in our experiments. Thus, the \textit{reg-DDPG} uses much more information than the \textit{prob-DDPG} approach.

A point forecast and a posterior distribution convey very different types of  information. A point prediction of $S_{t+1}$ provides one conditional estimate of the next signal, whereas the  posterior $\hat{\Phi}_t$ retains the uncertainty associated with the latent regime. For the partially observable control problem considered here, this uncertainty is relevant because different latent regimes imply different future dynamics and therefore different optimal decisions. Consequently, the posterior representation provides a more direct representation of the information required by the control problem when comparing to a single point forecast.}

\section{Results}
\label{sec:results-synthetic}

We now test the capabilities of the approaches outlined above on synthetic data and for increasing levels of complexity for the DGP as defined in Eq.~\eqref{eqn:OU-SDE}. To this end, as anticipated, we test our algorithm on environments where just $\theta_t$ is a Markov chain, where $\theta_t, \kappa_t$ are independent Markov chains, and finally, for the most complete case, where $\theta_t, \kappa_t$ and $\sigma_t$ are modelled via independent Markov chains. We employ both the two-step approaches and the one-step approach over a set of testing episodes, and subsequently compare the results, in terms of cumulative rewards, obtained by employing the trading strategies found by the DDPG agent.

In the testing phase, we compute the total reward for each of the $M$ testing episodes as $R_n = \sum^n_{t = 1} r_t$ where $n$ is the number of time steps in each of the $M$ testing episodes and $r_t$ is as defined in Eq.~\eqref{eqn:reward}. For each experiment, we report the average and the standard deviation of the cumulative rewards over the $M$ test iterations. The method that shows a higher average reward is the one that delivers the best {{approximated}\footnote{There is no general guarantee that the algorithms have found the globally best policy.} optimal trading policy for the analysed problem. {{We set the discounting rate $\gamma = 0.999$.} The simulation parameters for each experiment are reported in Table~\ref{tab:sim_pmts}. The transition rate matrices for the time varying parameters are 
\[
A_{\theta} = \begin{bmatrix}
  -0.1&  0.05&  0.05\\ 
  0.05& -0.1 &  0.05\\
  0.05&  0.05& -0.1\\
\end{bmatrix}
,\qquad 
A_{\kappa} = \begin{bmatrix}
  -0.1& 0.1  \\
  0.1&  -0.1 \\
\end{bmatrix},
\qquad \text{and} \quad
A_{\sigma} = \begin{bmatrix}
  -0.1& 0.1  \\
  0.1&  -0.1 \\
\end{bmatrix}.
\]
{{We have chosen to use symmetric transitions so that no one regime is preferred over another. The parameters used in the DGP simulations and in the algorithms are reported in Tables~\ref{tab:shared_pmts} and \ref{tab:sim_pmts}.}
\subsection{$\theta_t$ is a Markov chain}
{The first experiment considers the simplest case for the trading problem, i.e. where the mean-reversion level $\theta_t \in \{ \phi_1 = 0.9, \phi_2 = 1, \phi_3 = 1.1\}$ evolves as a Markov chain and $\kappa$ and $\sigma$ are held constant with values reported in Table~\ref{tab:sim_pmts}. In this setting, we first introduce two benchmarks that either consider no additional information on the signal process or feature the true regime value, we then turn to compare the results obtained from these two benchmarks to those obtained by the three deep RL algorithms introduced above: \textit{reg-DDPG, prob-DDPG and hid-DDPG}.
}

\begin{wraptable}{r}{0.45\textwidth}

    \centering
    \caption{{Information supplied to the DDPG actor for the five approaches considered in the experiments.}}
    \label{tab:information_features}
    \begin{tabular}{l l}
        \toprule
        Method & Input states \\
        \midrule
        \textit{na\"ive-DDPG}  & $(S_t,I_t)$ \\
        \textit{oracle-DDPG} & $(S_t,I_t,\theta_t)$ \\
        \textit{reg-DDPG}   & $(S_t,I_t,\hat S_{t+1})$ \\
        \textit{prob-DDPG}  & $(S_t,I_t,\hat{\Phi}_t)$ \\
        \textit{hid-DDPG}   & $(S_t,I_t,h_t)$ \\
        \bottomrule
    \end{tabular}
\end{wraptable}
The first benchmark is the \textit{na\"ive-DDPG}, which has access only to the observable variables $(S_t,I_t)$ and therefore does not receive any explicit or implicit information about the latent regime. At the opposite extreme, the \textit{oracle-DDPG} has direct access to the true value of $\theta_t$ at each time step; i.e., has access to a state consisting of $(S_t,I_t, \theta_t)$. The five approaches therefore differ only in the way information about the latent regime is made available and used by the actor while maintaining the same architecture reported in Table~\ref{tab:shared_pmts}, the different features used in the five approaches are summarized in Table~\ref{tab:information_features}.

{Comparing these approaches allows us to distinguish the effect of the \textit{information} available to the agent from the particular mechanism used to represent that information. The \textit{oracle-DDPG} provides an (unrealistic) upper-bound benchmark based on perfect knowledge of the latent regime, whereas the \textit{na\"ive-DDPG} represents the setting in which no additional information about the latent state is provided.  We report this ablation study in the simplest environment, where only $\theta_t$ switches among three regimes. This environment is the \emph{simplest environment for the the na\"ive agent}, as it contains the least latent information and it also provides easily interpretable results.
The remaining three approaches investigate what information about the latent regime best informs trading.}

{
The average reward for the (infeasible) \textit{oracle-DDPG} over $500$ testing iterations is $110.17$ with a standard deviation of $8.90$, whereas the \textit{na\"ive}-DDPG approach scores an average reward of $8.82$ with a standard deviation of $8.76$. The large performance gap highlights the value of access to the true latent regime. 
Recall the \textit{oracle-DDPG} can condition its policy on the hidden state whereas the \textit{na\"ve-DDPG} relies only on observable variables.
\begin{figure}[H]
    \centering
    \resizebox{\textwidth}{!}{
    \begin{tabular}{cc}
    \textit{oracle-DDPG}
    & \textit{na\"ive-DDPG}
    \\
    \includegraphics[height=0.22\textheight]{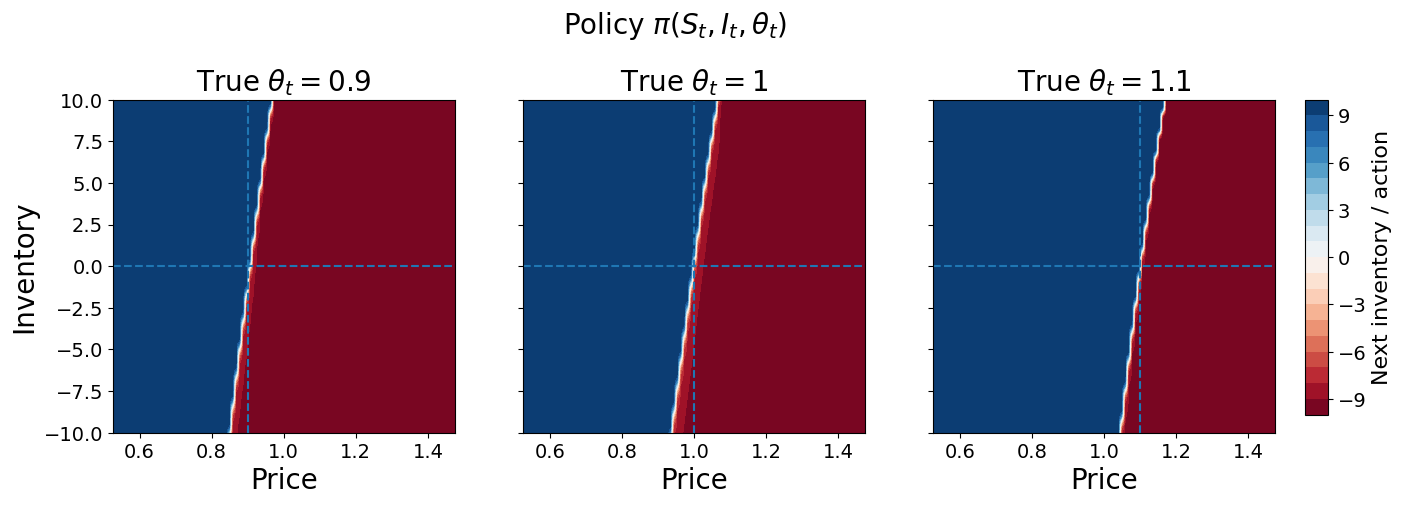}
    &
    \includegraphics[height=0.2\textheight]{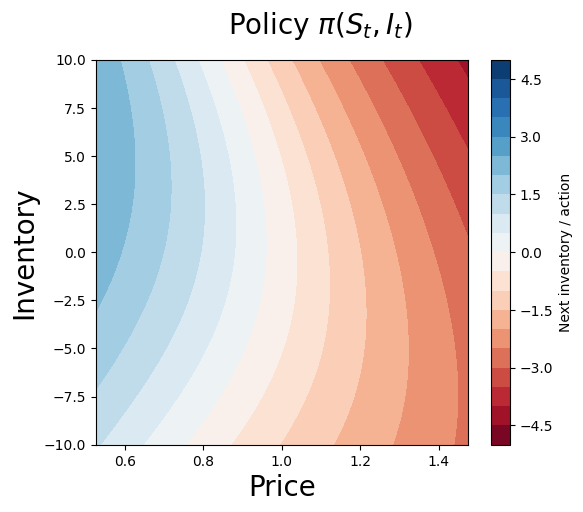}
    \end{tabular}}
    \caption{{Policy learned by the \textit{oracle-DDPG} (left three) and \textit{na\"ive-DDPG } (right) as a
    function of the price signal $S_t$ (horizontal axis) and the inventory $I_t$ (vertical axis), for each of the three true regimes
    $\theta_t \in \{0.9, 1.0, 1.1\}$ (left to right for \textit{oracle-DDPG}). Colour encodes the target  inventory, i.e.\ the action: blue for long positions, red for short positions, and white for no-trade region. The vertical dashed line marks the mean-reversion level $\theta_t$ of each panel, the horizontal dashed line marks zero inventory for the \textit{oracle-DDPG}, moreover the buy/sell threshold shifts to the right as $\theta_t$ increases.
    }}
    \label{fig:heatmap_pols_full}
\end{figure}
Fig.~\ref{fig:heatmap_pols_full}  shows the estimated policy heat-maps as a function of inventory, asset price, and regime (for the \textit{na\"ive-DDPG} there is no dependence on regime). The \textit{oracle-DDPG} effectively learns distinct policies for the different regimes and has sharp boundaries between going long or short as it exploits knowledge of the regime.  The optimal policies also have small no-trade bands (the white area) due to transaction costs. Contrastingly, in the \textit{na\"ive-DDPG} optimal policy there is no sensitivity to regime, as it has no knowledge of it nor does it infer it. Further, while there is some sensitivity to price, there is less sensitivity to inventory levels.  Although the resulting policy retains some dependence on the observable signal and the inventory,  as expected, it cannot exploit regime-specific mean-reversion information.}

{{Having established the value of knowing the true regime, we next investigate how much of this information can be recovered through the different learning mechanisms. We do so by comparing cumulative rewards of the three methods: \textit{reg-DDPG, prob-DDPG, and hid-DDPG}. Summary statistics (the mean, standard deviation, and left $CVaR_{10\%}$) of the results are summarised in Table~\ref{tab:results_simulation}, the distributions are shown in Fig.~\ref{fig:theta_rew}, and the learned policies for the three algorithms are shown in Fig.\ref{fig:hid_ddpg},\ref{fig:reg_ddpg} and Fig.~\ref{fig:prob_ddpg}.}
{
\begin{table}
    \centering
    \caption{{Average cumulative rewards, their standard deviation, and the $10\%$ left-tail CVaR after $n = 2{,}000$ trades from the $M=500$ test episodes of trading, for each approach and each DPG. The CVaR is the mean cumulative reward over the worst $10\%$ of episodes, $\mathbb{E}[R \mid R \le q_{10\%}]$.}}
    \resizebox{\textwidth}{!}{
    
    \begin{tabular}{
    l c
    S[table-format=2.2] S[table-format=1.2] S[table-format=2.2]
    c
    S[table-format=2.2] S[table-format=1.2] S[table-format=1.2]
    c
    S[table-format=-1.2] S[table-format=1.2] S[table-format=-1.2]}
    \toprule
         \multirow{2}{*}{Model rewards} & & \multicolumn{3}{c}{$\theta_t$}
         & & \multicolumn{3}{c}{$\theta_t,\,\kappa_t$}
         & & \multicolumn{3}{c}{$\theta_t,\,\kappa_t,\,\sigma_t$} \\
         \cmidrule{3-5}\cmidrule{7-9}\cmidrule{11-13}
           & & {Ave.} & {Std.} & {CVaR$_{10\%}$}
           & & {Ave.} & {Std.} & {CVaR$_{10\%}$}
           & & {Ave.} & {Std.} & {CVaR$_{10\%}$} \\
         \midrule
        \textit{hid-DDPG}  & 
        & 15.70 & 1.39 & 13.29 & 
        & 8.03  & 1.54 & 5.55  & 
        & 1.28  & 3.49 & -5.08 \\
        \textit{reg-DDPG}  & 
        & 8.69  & 0.60 & 7.67  & 
        & 2.95  & 0.30 & 2.42  & 
        & -0.08 & 0.11 & -0.27 \\
        \textit{prob-DDPG} & 
        & 25.65 & 3.35 & 19.49 & 
        & 15.59 & 3.83 & 8.78  & 
        & 4.52  & 3.75 & -1.62 \\
         \bottomrule
    \end{tabular}}
    \label{tab:results_simulation}
\end{table}}
\begin{figure}[H]
\centering
\includegraphics[width=1\textwidth, angle = 0]{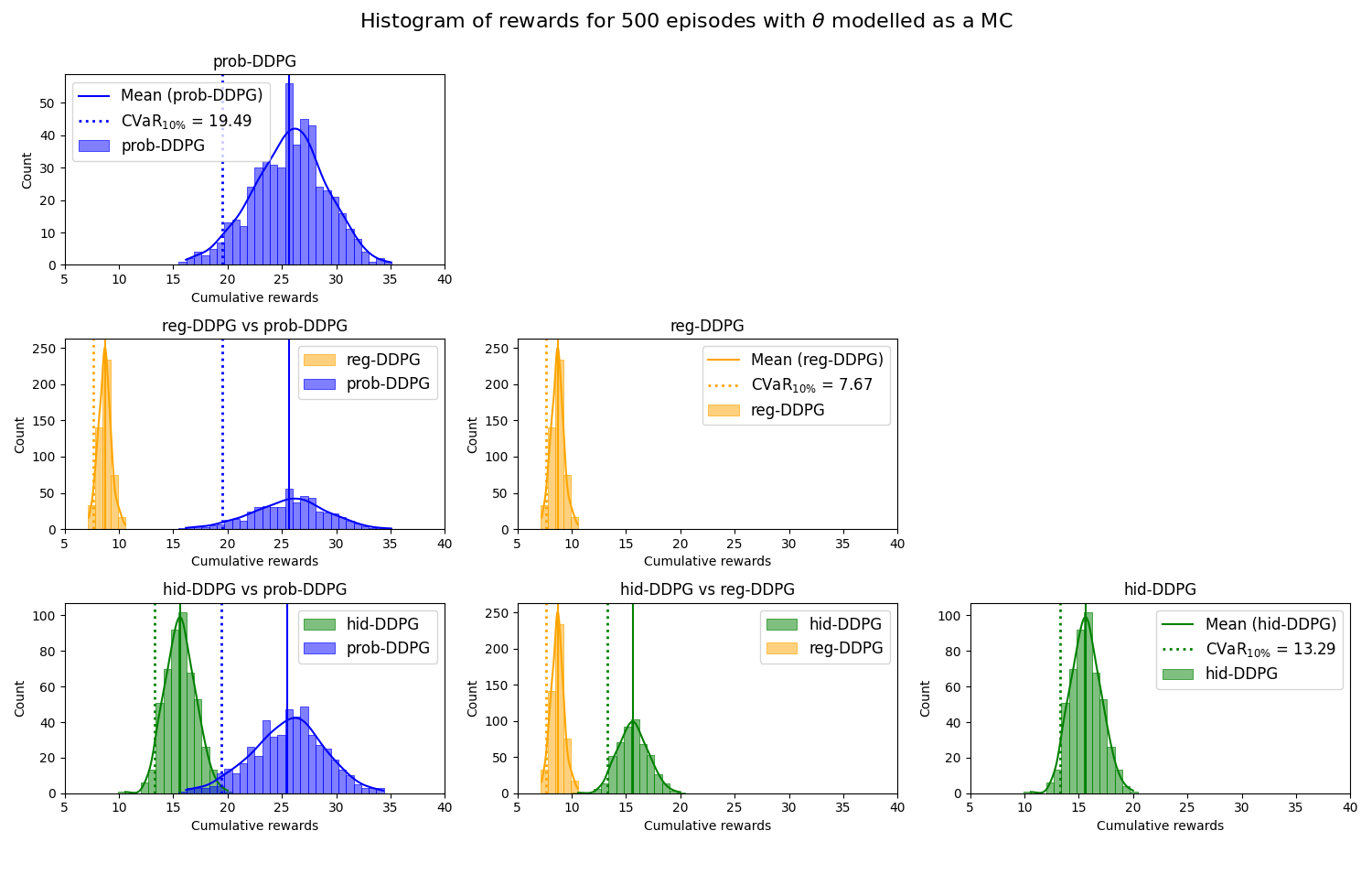}
    \caption{Comparison of the cumulative rewards for the different DDPG approaches when $\theta_t$ is a Markov chain.}
    \label{fig:theta_rew}
\end{figure}

{As reported in Table~\ref{tab:results_simulation} and  Fig.~\ref{fig:theta_rew}, the \textit{prob-DDPG} approach achieves the highest average cumulative reward and $CVaR_{10\%}$ among the three learning based algorithms analysed. Moreover, all three approaches achieve non-negative cumulative rewards across the simulations. 
In the \textit{hid-DDPG} approach, the hidden state of the GRU is learned jointly with the DDPG policy and is fed directly to the Actor network as part of its input state. The last layer of hidden states of the GRU therefore appears to carry  informative about the latent state of the signal $S_t$ and facilitates  policy learning. However, because the latent representation is learned jointly with the control objective it is not possible to decouple it and attach probabilistic, predictive, or economic interpretation to it.  Looking at the optimal strategy across simulations reported in the left most panel of Fig.~\ref{fig:hid_ddpg}, due to the clustering of inventory, it appears that the policy has a preference for a limited set of trading decisions. Some of these actions appear counter-intuitive. For instance, the policy recommends selling when $S_t$ is relatively low and $I_t$ is positive, or buying when $S_t$ is relatively high and $I_t$ is low. However, such decisions do not stem solely from the current state $(S_t,I_t)$, because the GRU hidden state contains some information about the latent regime. This makes the resulting policy substantially less interpretable than a belief-based policy.

The \textit{prob-DDPG} approach instead provides the actor with the posterior probability vector $\hat{\Phi}_t$ estimated by the GRU. Unlike the hidden-state representation, this information has a direct probabilistic interpretation and describes the agent's uncertainty about the current latent regime. As shown in the left panel of Fig.~\ref{fig:prob_ddpg}, the resulting policy exhibits a clear dependence on the signal and inventory. Buy actions are predominantly concentrated for relatively low values of $S_t$, while sell actions become more frequent for relatively high values of $S_t$. At the same time, the policy can condition these actions on the estimated regime probabilities.  This behaviour is even better illustrated in the first row of Fig.~\ref{fig:scatters_thetas}, where actions are shown as a function of $S_t$ and $I_t$ conditional on the estimated posterior distribution. The shades indicate the probability assigned by the GRU to the different mean-reversion regimes, while the colour of the point boundary indicates whether the corresponding inventory rebalancing $q_t$ is a buy or a sell. The posterior therefore allows the same observable state $(S_t,I_t)$ to lead to different actions depending on the inferred latent regime.  For example, when the signal is relatively low, a high posterior probability of the regime $\theta_t=0.9$ indicates that the signal is expected to revert towards a lower long-run level. In this situation, an agent that is already long may prefer to reduce her position rather than increase it further. Conversely, when the inferred regime corresponds to a higher mean-reversion level, the same low value of $S_t$ can provide an incentive to increase the position in anticipation of upward mean reversion. The posterior representation therefore allows the policy to condition its trading decisions jointly on the current signal, inventory, and inferred latent regime. 

In the \textit{reg-DDPG} approach, the GRU is trained to forecast the next value of the observable signal, $\hat S_{t+1}$, and this forecast is subsequently used by the DDPG agent. The policy reported in the left panel of Fig.~\ref{fig:reg_ddpg} appears less concentrated around a small number of actions than the \textit{hid-DDPG} policy. Its buy and sell decisions are also more directly related to combinations of the current signal and inventory. For example, when the signal is low and the current inventory is relatively small, the forecast of the next signal can provide an incentive to buy in anticipation of upward movements.  

The comparison between the three approaches highlights an important distinction between the representations learned by the recurrent network. The \textit{hid-DDPG} approach learns an unconstrained hidden representation jointly with the control policy, whereas \textit{reg-DDPG} learns a point forecast of the next observable signal. The \textit{prob-DDPG} approach instead explicitly learns a probability distribution over the latent regimes. The latter representation preserves information about uncertainty that is discarded by a point forecast
and provides the DDPG agent with information that is directly related to the latent dynamics.  Finally, comparing the three learning-based approaches with the two benchmark cases provides an empirical assessment of the value of latent state information. The (infeasible) oracle-DDPG achieves an average reward of $110.17$, while \textit{prob-DDPG} reaches $25.65$, corresponding to approximately $23\%$ of the perfect-information benchmark. 

The belief-based \textit{prob-DDPG} approach consistently outperforms both alternative learned representations and the \textit{na\"ive-DDPG} benchmark. These results show that latent-regime information can be extracted from the observable signal and that representing it explicitly as a belief distribution can meaningfully improve trading performance over alternative filtering and learning approaches. The \textit{oracle-DDPG} quantifies the value of latent information, while the \textit{na\"ive-DDPG} captures the cost of ignoring it. The three learned approaches show how this information can be represented and exploited, with posterior-based learning producing quantitatively and qualitatively better policies.  }  

\subsection{$\theta_t,\,\, \kappa_t$ are Markov chains}

\begin{figure}[H]
\centering
\includegraphics[width=1\textwidth, angle = 0]{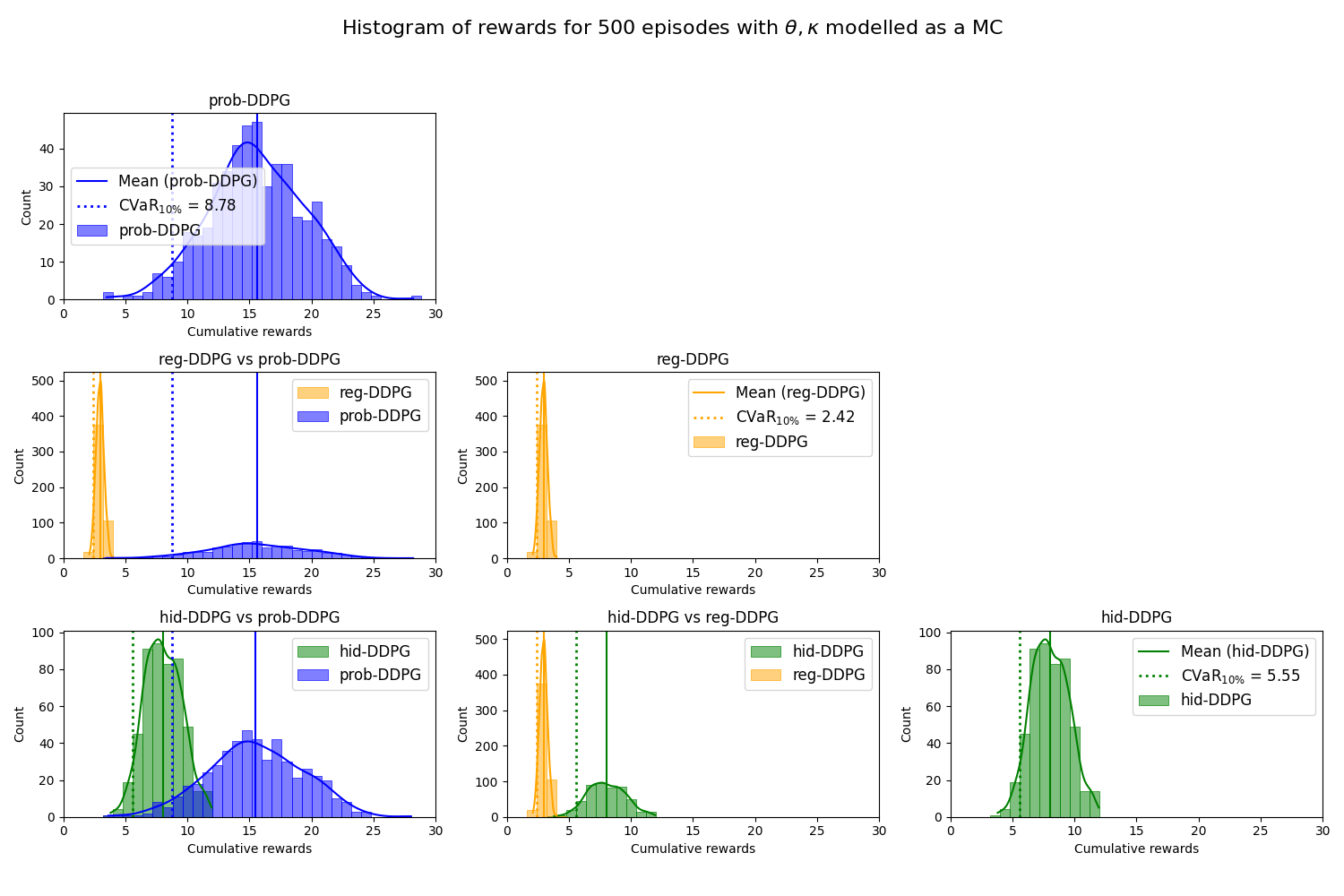}
    \caption{Comparison of the cumulative rewards for the different DDPG approaches when $\theta_t$ and $\kappa_t$ follow a  Markov chain.}
    \label{fig:kappa_rew}
\end{figure}

We next consider the case when both the level $\theta_t$ and the speed of mean reversion $\kappa_t$ in Eq.~\eqref{eqn:OU-SDE} are modelled as independent Markov chains. As reported in Fig.~\ref{fig:kappa_rew}, in this setting, once again all three learning approaches \textit{reg-DDPG}, \textit{hid-DDPG}, and \textit{prob-DDPG} produce non-negative rewards across the simulations. As seen in Table~\ref{tab:results_simulation}, in this setting the  average rewards are lower than in the setting with only $\theta_t$ being regime dependent. This is to  be expected as  the parameter dynamics is more complicated and more difficult to learn. 
As before,  the \textit{reg-DDPG} approach produces rewards that are generally better than the \textit{hid-DDPG} approach. This is again due to the more complex signal dynamics, with the time-varying $\kappa_t$ making the next signal value even more difficult to predict.
The \textit{prob-DDPG} approach once again outperforms the two others, both in expectation and in tail risk, although the rewards are generally lower than \textit{prob-DDPG} performance in the case when only $\theta_t$ shifts regimes. 

We next analyse the actions chosen by the Actor network per level of inventory $I_t$ and trading signal $S_t$. To this end, the middle panels of Fig.~\ref{fig:actions_varie} show that have policies that vary considerably depending on the type of information learnt. Indeed, Figs.~\ref{fig:reg_ddpg} and \ref{fig:hid_ddpg} show that the actions of the \textit{reg-DDPG} and \textit{hid-DDPG} lead to inventory levels concentrated in a relatively narrow region of the admissible domain, $[I_{\min},I_{\max}]$. We conjecture that this limited use of the inventory range partly explains their lower rewards in Fig.~\ref{fig:kappa_rew}. In contrast, the \textit{prob-DDPG} policy in Fig.~\ref{fig:prob_ddpg} produces more dispersed inventory holdings, suggesting that explicitly providing posterior probabilities over the regimes governing the DGP's mean-reversion level allows the agent to exploit latent-regime information more effectively. Moreover, the resulting policy admits a natural economic interpretation. In fact, the second row of Fig.~\ref{fig:scatters_thetas} shows that the \textit{prob-DDPG} policy $q_t$ varies not only with the signal $S_t$ and inventory $I_t$, but also with the posterior probabilities of the mean-reversion regimes, now in the presence of two possible mean-reversion speeds $\kappa_t$. Although none of the learned approaches is explicitly provided with information about $\kappa_t$, conditioning on the posterior distribution of $\theta_t$ appears more effective than using either hidden states or forecasts of the next signal value. This is reflected both in the higher cumulative rewards and in the greater interpretability of the resulting policy.

\subsection{$\theta_t,\,\, \kappa_t$ and $\sigma_t$ are Markov chains}

\begin{figure}[h]
\centering
\includegraphics[width=1\textwidth, angle = 0]{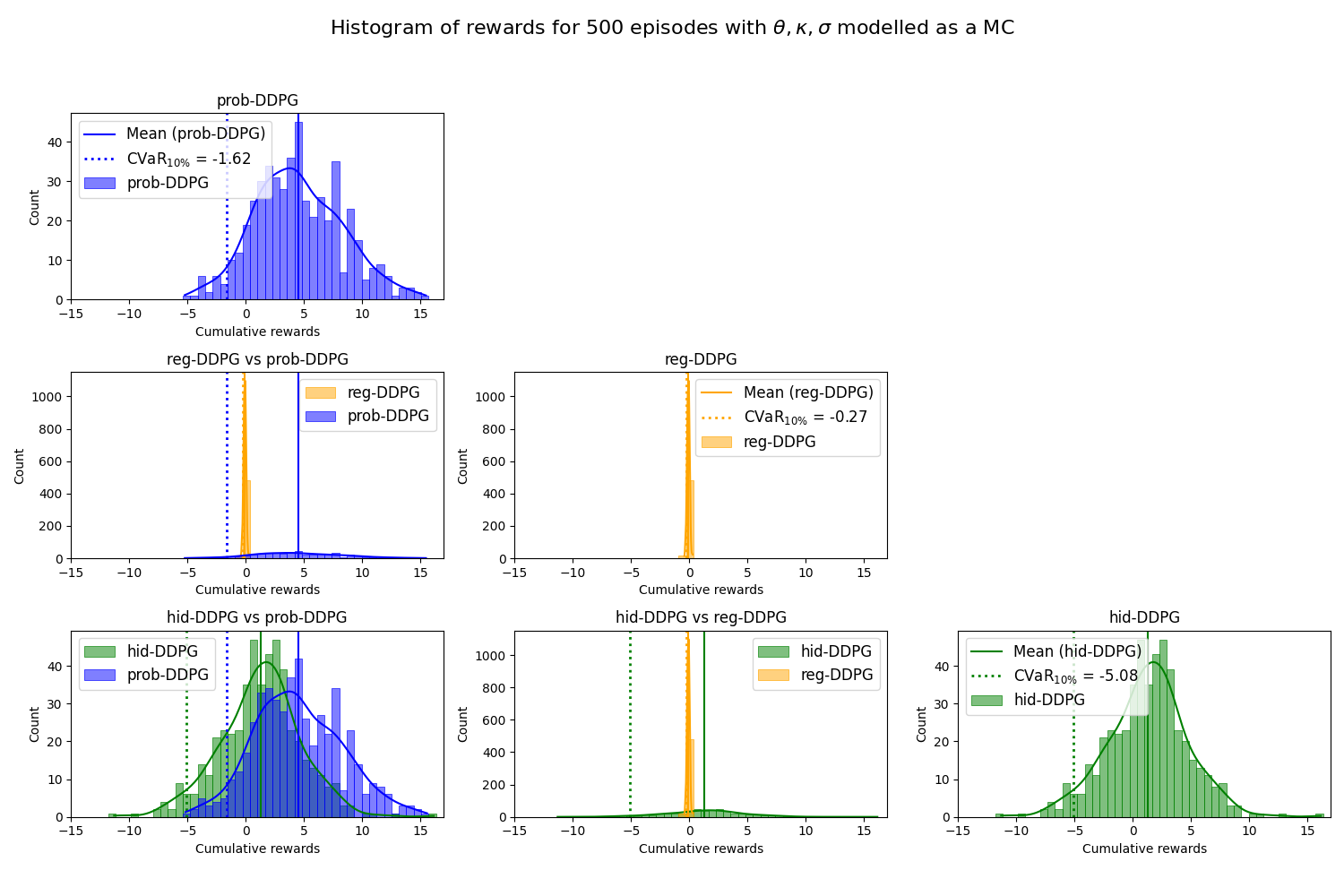}
    \caption{Comparison of the cumulative rewards for the different DDPG approaches when $\theta_t$, $\kappa_t$, and $\sigma_t$ follow a Markov chain.}
    \label{fig:sigma_rew}
\end{figure}

In this last and most complex setting, we allow volatility, mean-reversion speed, and mean-level to switch regimes independently according to the  parameters reported in Table~\ref{tab:sim_pmts}. 

The average cumulative rewards reported in Table~\ref{tab:results_simulation} once again shows that  the \textit{prob-DDPG} outperforms the other two approaches in expectation. This is the first experiment, however, where, the \textit{prob-DDPG} does not have the best tail risk performance. The \textit{reg-DDPG} outperforms it in this respect; however, the expected reward of \textit{reg-DDPG} is slightly negative (but essentially zero), and is effectively a do-nothing strategy, while \textit{prob-DDPG} has expected reward about 4.52. An investor who is risk averse, but also profit seeking would likely prefer the performance of the \textit{prob-DDPG} over that of the others.

We postulate that the poor performance of \textit{reg-DDPG} stems from the lower informative content that such an estimate provides this complex setting. As mentioned earlier, the policy is effectively a do-nothing one due to the difficulty in extracting good next step predictions from the signal.

As shown in Table~\ref{tab:results_simulation} and Fig.~\ref{fig:sigma_rew}, the \textit{hid-DDPG} outperforms the \textit{reg-DDPG} but underperforms the \textit{prob-DDPG} in terms of cumulative rewards. Figure~\ref{fig:hid_ddpg} suggests that the GRU hidden states remain informative in this more complex environment, although they produce larger trading actions and are less interpretable than posterior probabilities.

Turning to the \textit{prob-DDPG} approach\footnote{In this case, we set the look-back window to $W=20$.}, Table~\ref{tab:results_simulation} shows that average cumulative rewards decrease in this more complex environment, but remain higher than those of the other learned approaches, while their standard deviation remains relatively stable across the three experiments. These results again highlight the benefit of explicitly learning the latent regime. Moreover, Figs.~\ref{fig:scatters_thetas} and \ref{fig:prob_ddpg} show that the resulting policy is interpretable: buy/sell decisions depend not only on the current signal $S_t$ and inventory $I_t$, but also on the inferred $\theta_t$ regime.

\section{Pairs-trading application}
{Having examined the performance of the proposed approaches on synthetic data in Section~\ref{sec:results-synthetic}, we now evaluate them on real market data in the classical \textit{pairs-trading} setting. This analysis provides an empirical proof of concept, testing whether the proposed regime-aware, model-agnostic framework can be applied to real high-frequency data and exploit information about latent mean-reversion dynamics in a financially meaningful trading problem.

We consider a risk-neutral agent trading a portfolio of two co-integrated assets with fixed weights, with the objective of maximising cumulative trading revenues. We first estimate the portfolio composition using classical econometric tools (as in \cite{cartea2015algorithmic}) and derive an analytical model-based benchmark strategy. We then apply the model-agnostic deep RL approaches developed in the previous sections, which operate on the partially observed signal and infer information about its latent regime from the observed history. The resulting strategies are compared in terms of cumulative rewards over the testing period, allowing us to assess whether explicitly representing information about the latent mean-reversion regime improves the performance of the deep RL agent under real market conditions.

As in the synthetic experiments, we assume a regime-switching long-run level $\theta_t$, while keeping the mean-reversion speed $\kappa$ and volatility $\sigma$ constant. This provides a parsimonious specification while retaining the partially observable structure central to the proposed methods.
}

\subsection{Dataset, preliminary data analysis and market model}
\label{sec:preliminary}
We consider two assets traded on the NASDAQ between August 29, 2025 and September 5, 2025, namely the Intel (INTC) stock and the {{VanEck} semi-conductors ETF (SMH) whose holdings are around 20\% composed by INTC shares. 

We use high frequency observations of the limit order book for both the assets, downloaded from the LOBSTER database, we consider data up to the first level of the order book and we select only trade events, so that each element considered in the time series represents a transaction. We then calculate the mid-price and filter the mid-prices at a frequency of 1 second, such that we have a clean dataset that displays prices at even intervals of time. Both time series are displayed in Fig.~\ref{fig:SMHvsINTC}.
\begin{figure}[ht]
    \centering
    \includegraphics[width=0.8\linewidth]{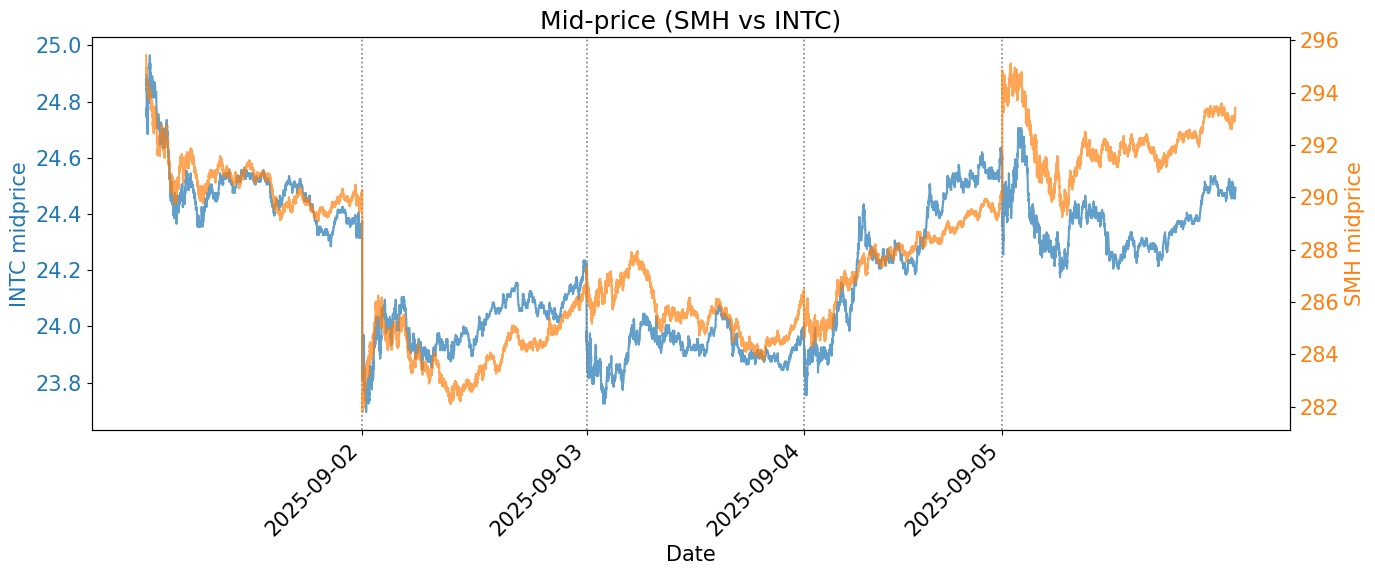}
    \caption{Mid-prices for SMH and INTC, prices are intended in US dollars.}
    \label{fig:SMHvsINTC}
\end{figure}
{{Firstly, to build the pairs-trading signal we use the Johansen test (\cite{johansen1988statistical, johansen1991estimation}), and record the results in Table~\ref{tab:johansen}. The test confirms the presence of at least one stationary  linear combination of the two mid-prices and we denote the linear combination by 
\begin{equation}   \label{eqn:cointegration}
\tilde S_t = \alpha_1 \,S^{\mathrm{smh}}_t + \alpha_2 \,S^{\mathrm{intc}}_t.    
\end{equation} 
\begin{table}[H]
\centering
\caption{Johansen Co-integration Test (5\% level). Trace statistic, maximum Eigenvalue statistic and their critical values of the Johansen test for co-integration. The H0 is no co-integrating vector.}
\begin{tabular}{c ccc ccc}
\toprule
$r$ & Trace Stat & Crit & Reject & Max-Eig Stat & Crit & Reject \\
\midrule
0 & 27.456 & 18.399 & Yes & 19.222 & 17.148 & Yes \\
\bottomrule
\end{tabular}
\label{tab:johansen}
\end{table}
Our signal is, therefore, a scalar process obtained by holding the two assets in fixed proportions $\alpha_1$, $\alpha_2$.}

{\paragraph{Estimating the co-integrating weights.} To recover the co-integrating weights $(\alpha_1, \alpha_2)$, we follow the procedure outlined in \cite{cartea2015algorithmic}, therefore we start by fitting a bivariate vector autoregression (VAR) model to the mid-prices of the two assets considered. Let $\mathbf{S}_t = (S^{\text{smh}}_t, S^{\text{intc}}_t)^\top  \in \R^2$, the VAR model is given by 
$$
\mathbf{S}_t= \vec{C} + B\mathbf{S}_{t-1} + \vec{\epsilon}_t
$$
where $\vec C\in\mathbb{R}^2$ is a constant vector, $B\in\mathbb{R}^{2\times2}$ is the matrix of autoregressive coefficients, and $\vec\epsilon_t\in\mathbb{R}^2$ is a vector white noise. The estimated coefficients are reported in Table~\ref{tab:var_model}. From the VAR estimates we obtain the matrix of mean-reversion coefficients and the corresponding vector of long-run levels (with $\Delta t=1$): 
$$
K = \frac{\mathbb{I}-B}{\Delta t} =
\begin{bmatrix} 1.40\cdot 10^{-4} & -1.52\cdot 10^{-5}\\ -7.78\cdot 10^{-4} & 3.05\cdot 10^{-4}
\end{bmatrix}, \qquad 
\vec\Theta = K^{-1}\vec C\,\Delta t = 
\begin{bmatrix} 287.77\\ 24.183\end{bmatrix}.
$$
Then, we reduce this system to a portfolio with scalar dynamics and use it for trading. 
Firstly, diagonalising $K = U \Lambda U^{-1}$, we obtain eigenvalues $\Lambda_1=8.67\cdot10^{-5},
\Lambda_2=3.59\cdot10^{-4}.$ Then, we select the portfolio that is the most exposed to mean reversion by projecting the price vector onto the eigenbasis and selecting the direction associated with the highest eigenvalue. Therefore, we find that the weights are $(2.856,-0.804)$. Our co-integrated scalar portfolio is:
\begin{equation}
    \label{eqn:TildeS}
    \tilS = 2.856\, S^{\text{smh}}_t- 0.804\,S^{\text{intc}}_t 
\end{equation}
Along the selected direction the dynamics can be represented by an Ornstein-Uhlenbeck SDE as in Eq.~\eqref{eqn:OU-SDE}, which in discrete time corresponds to the AR(1) representation
\begin{equation}
    \tilde S_t=a+b\tilde S_{t-1}+\tilde\eta_t,
\label{eqn:recursion_ou}
\end{equation}
where $b=e^{-\tilde\kappa\Delta t},
a=\tilde\theta(1-b),
\tilde\theta=u^\top\vec\Theta,$ with $u$ denoting the selected eigenvector direction and $\tilde\eta_t$ the corresponding innovation term.
We divide the resulting series into a training set, containing observations up to September 4, 2025 and a testing set corresponding to September 5, 2025 (Figs.~\ref{fig:S_tilde_regimes}--\ref{fig:S_tilde_test}). Thus, the agent is trained on almost one week of data and its trading performance is evaluated on the following day.
We then apply min--max normalisation using only the training set. This provides a common scale for the subsequent regime estimation and for the learning algorithms.
The normalised series is characterised by the AR(1) coefficient $b = 0.99964,\, \tilde\theta=0.44$.
}
\begin{figure}[H]
\centering
\includegraphics[width=0.8\linewidth]{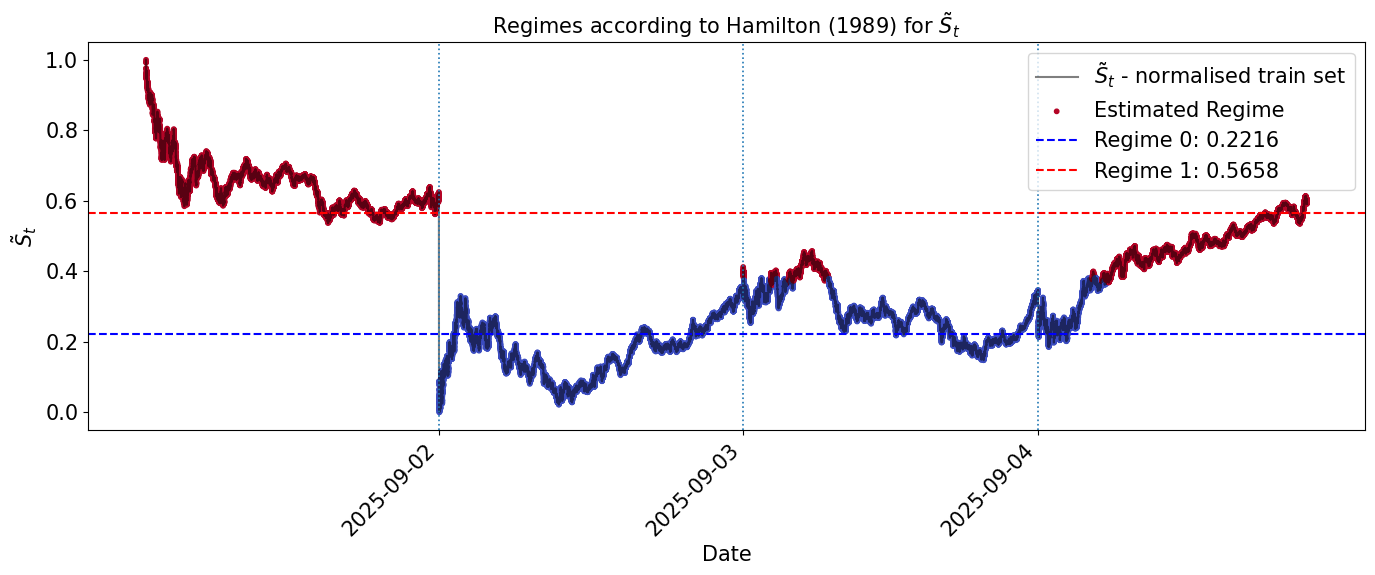}
\caption{Normalised time series for the training phase of the co-integrated portfolio $\tilde{S}_t$ and the most probable regimes at each time step.}
\label{fig:S_tilde_regimes}
\end{figure}
\begin{figure}
\centering
\includegraphics[width=0.8\linewidth]{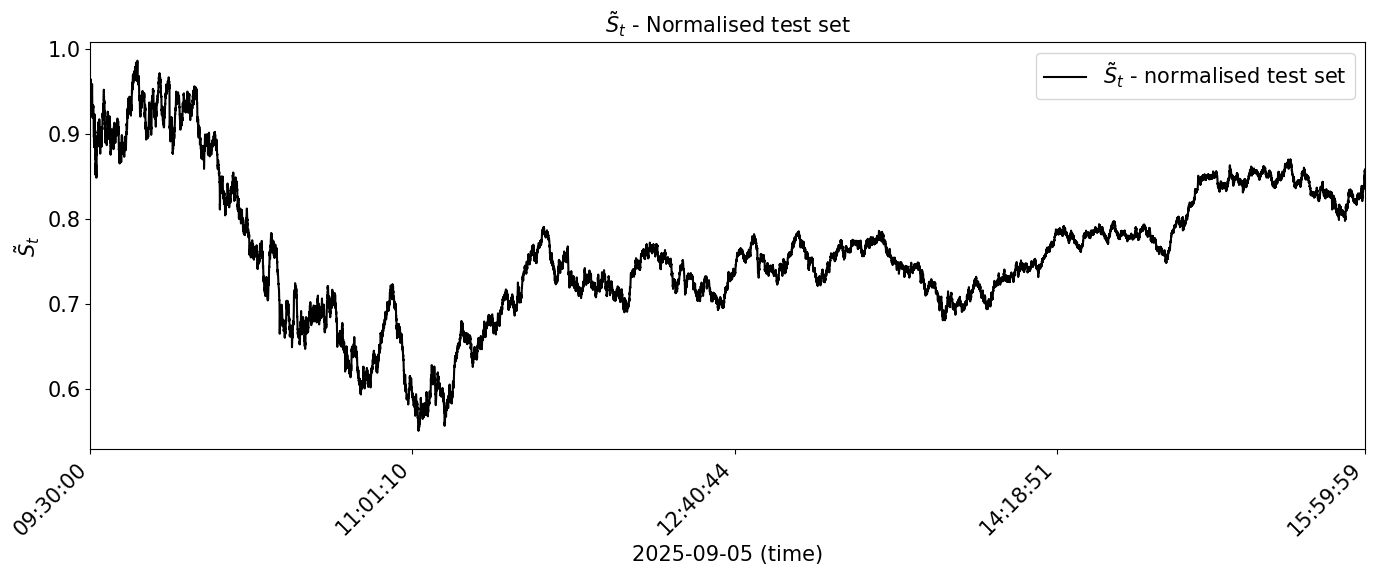}
\caption{Normalised time series for the testing phase of the co-integrated portfolio $\tilde{S}_t$.}
\label{fig:S_tilde_test}
\end{figure}
{\paragraph{Regime estimation.} 
We now turn to test for the presence of possible regimes in the (normalised) co-integrated portfolio and estimate their values. Following the procedure outlined in \cite{hamilton1989}, Chapter 22; we find two regimes for $\tilS$, more specifically we estimate $\widehat{\theta}_1 = 0.2216, \widehat{\theta}_2=0.5658$. Figure~\ref{fig:S_tilde_regimes} reports the most probable regime at each time step. 

To assess whether the two estimated regimes correspond to statistically distinct long-run levels, we test $
H_0:\theta_1=\theta_2
\,\,\text{against}\,\,
H_1:\theta_1\neq\theta_2.
$
The test statistic is

$$
z=
\frac{\widehat{\theta}_2-\widehat{\theta}_1}
{\sqrt{
SE(\widehat{\theta}_1)^2+
SE(\widehat{\theta}_2)^2-
2\,\mathrm{Cov}(\widehat{\theta}_1,\widehat{\theta}_2)
}},
$$
with corresponding two-sided $p$-value
$
p=2\big(1-\Psi(|z|)\big),
$
where $\Psi$ denotes the cumulative distribution function of a standard normal random variable. We obtain $z=421$ and $p<0.001$, providing strong evidence that the two estimated regime levels are statistically different.
The estimated transition matrix is
$
\hat A=
\begin{bmatrix}
0.9998&0.0002\\
0.0002&0.9998
\end{bmatrix}.
$
The diagonal elements indicate highly persistent regimes, while the small off-diagonal probabilities imply that transitions between regimes are relatively infrequent. When transitions occur, however, the change in the estimated long-run level is sharp, as also illustrated in Fig.~\ref{fig:S_tilde_regimes}.

Given the high persistence of the regimes, we use a longer history window $W$ than in the synthetic experiments to improve the estimation of the agents' beliefs. At the same time, $W$ is kept relatively short so that, during training, randomly sampled batches expose the algorithms to a larger number of distinct segments of the signal. This provides a compromise between accurate belief estimation and sufficient variability in the training samples.\footnote{A higher number of regimes, i.e., $k>2$, would lead to the presence of absorbing states. We suspect that this is due to the limited amount of data used in this estimation.} 
}

\subsection{Trading experiments}
{%
Having constructed the traded signal and estimated its latent regime structure, we now compare both a model-based benchmark rule and a rule-based trading strategy with the deep RL approaches developed in the previous sections. The same transaction-cost specification introduced in Section~\ref{sec:model} is used throughout the trading experiments. In the test time-series, September 5, we chose not to consider the first and the last half-hour, i.e. after market opening and before closing, as these are the most volatile time windows.}
\subsubsection{Benchmark trading strategies}
\label{sec:plugin}
{
We first obtain a model-based (plug-in) benchmark trading strategy, using the AR(1) dynamics for the co-integrated portfolio and the parameters obtained from the data.
In this setting, the conditional expectation of the next signal value  is $\E_{t-1}\left[\tilS\right] = a + b\tilSminus$. Consequently, the model-implied expected one-step change in the traded signal is $m(\tilS) = (1-b)(\tilde{\theta} - \tilSminus)$. With  reward function $r_t$ in Eq.~\eqref{eqn:reward}, the optimal inventory to hold depends on the magnitude of the expected change in $\tilS$ and this expected change should more than compensate the transaction costs, which we assume is proportional to the inventory change.   
Therefore, conditional on the current information, the expected one-step reward is
\begin{equation}
    \begin{split}
        \E_{t-1}[r_t] =
    I_t\,\E_{t-1}[\tilde S_t-\tilde S_{t-1}]
    -
    \lambda |I_t-I_{t-1}| 
    \approx
    I_t\,m(\tilS)
    -
    \lambda |I_t-I_{t-1}|.
    \label{eqn:plugin_expected_reward} 
    \end{split}
\end{equation}
As the trader's actions \textit{do not} impact the signal's value and the trading decision is made at time $t$, in this model-based decision problem we use a \textit{one-step myopic} criterion and the trader seeks
\begin{equation}
     \tag{P2}
     \label{eqn:prob_2}
     \max_{I_t \in \left[I_{\text{min}}, I_{\text{max}}\right]} \left\{ I_t\,m(\tilS) - \lambda|I_t-I_{t-1}|\right\}.
\end{equation}
This criterion admits an explicitly solution in terms of  three possible trading regions. First, if, at time $t$, the agent chooses $I_{t}>I_{t-1}$ then the objective function in Eq.~\eqref{eqn:prob_2} becomes $ I_t(m(\tilS) - \lambda)) + \lambda I_{t-1}$ and, hence, buying is profitable only when $m(\tilS)>\lambda$. Second, similarly, selling is profitable only when $m(\tilS)<-\lambda$. 
Third, whenever $-\lambda < m(\tilS) < \lambda$, neither buying nor selling compensates the trader enough to cover the transaction cost $\lambda$, therefore in this region it is optimal to keep the current inventory fixed: $I_t = I_{t-1}$.

Combining the three cases, the plug-in model optimal strategy is
\begin{equation}
\label{eqn:plug_in}
I^*_t = 
    \begin{cases}
        I_{\max} \quad \text{if}\quad  (\tilde\theta-\tilSminus)> \lambda/(1-b)
        ,\\
        
        I_{t-1}\quad \text{if}\quad |\tilde\theta-\tilSminus|\leq \lambda/(1-b),
        \\
        I_{\min} \quad \text{if}\quad (\tilde\theta-\tilSminus)< -\lambda/(1-b)\,.
    \end{cases}
\end{equation}
}
{
The no-trading region above, may be expressed as the interval $\tilde{S}_t\in[\underline{S},\overline{S}]$, where
\begin{equation}
    \label{eqn:regions}
        \underline S
    =
    \tilde\theta-\frac{\lambda}{1-b},
    \qquad
    \overline S
    =
    \tilde\theta+\frac{\lambda}{1-b},
\end{equation}

In addition to this model-based rule, we also compare against a purely rule-based benchmark, the rolling $Z$-Score strategy. We chose to employ this additional benchmark as it is not model-based. Furthermore, inventory adjusts with the rolling average and standard deviation of the signal over a time window, and makes no assumption on either the presence of multiple regimes or their value (see \cite{chan2013algorithmic}). Following this trading rule, the agent accumulates inventory $I^{(W)}_t$ that is proportional to the negative rolling $Z$-Score of the portfolio itself, i.e.,
\begin{equation}
    I^{(W)}_t = - \frac{\tilde{S}_t - \mu^{(W)} }{{\zeta^{(W)}}}
\end{equation}
where $\zeta^{(W)}$ and $\mu^{(W)}$ are respectively the standard deviation and the average of the process over the previous $t-W$ observations.}
For the rolling $Z$-Score computation we use the same window $W$ used for other algorithms in this section, and we assume the same inventory levels constraints as for the other cases studied, clipping to either $I_{\max}$ or $I_{\min}$ the inventory suggested by the strategy.
\subsubsection{Trading with deep reinforcement learning}
Based on the results on synthetic data we now employ the approaches that performed the best in the case of $\theta_t,\,\kappa_t\,\,\sigma_t$ modelled as Markov chains, as that was the most complex environment. We thus tackle the pairs-trading problem with both the \textit{hid-DDPG} approach and the \textit{prob-DDPG} approach. The parameters used for the simulations are reported in Table~\ref{tab:real_pmts} for the DDPG part of the model,\footnote{The transaction cost $\lambda$ is smaller than in the synthetic experiments because it is scaled to the typical one-step movement of the traded signal.} while for the GRU part of the algorithm we keep the same hyperparameters as in the case of $\theta_t,\,\kappa_t\,\sigma_t$ modelled as Markov chains as reported in Table~\ref{tab:architectures}. 
During training the $\{\tilde{ S}_{u}\}^{t-W}_{u=t}$ are fed to the GRU and the inputs to both the Actor and Critic networks are randomly picked from the training set and have, as in the simulated case above, a length $W$. 
\begin{table}[H]
    \centering
    \begin{tabular}[\textwidth]{|c|c|c|}
        \hline
        \multicolumn{3}{|c|}{DDPG parameters} \\
        \hline
        train episodes $10,000$ & learning rate $= 0.001$ & batch $= 64$ \\
        \hline
        $\gamma = 0.999$ & $I_{\text{max}} = 10$ & $\lambda = 0.001$ \\
        \hline
        $\theta = \{0.2216, 0.5658\}$ & $W = 500$ & layers $= 6$ \\
        \hline
        hidden nodes $= 64$ & train observations $= 72,700$ & test observations $= 16,300$ \\
        \hline
    \end{tabular}
    \caption{Parameters for the DDPG algorithm.}
    \label{tab:real_pmts}
\end{table}

{{For this set of experiments we choose $W= 500$, and as we observe mid-prices every second, corresponds to about 8.3 minutes of observations.  As before, we set $\gamma = 0.999$, which corresponds to optimising over an interval of $1/(1-\gamma) = 1,000$ seconds or about $17$ minutes. The mean-reversion half-life of the traded spread is given by $t_{1/2}=\log 2/\tilde\kappa$, and with an estimated $\tilde\kappa=3.59\times10^{-4}\,\mathrm{s}^{-1}$, this corresponds to a half-life of the cointegration factor of about $32$ minutes. Thus, over a window of length $W = 500$ seconds the regime is essentially fixed. 
Drawing batches of random windows while training the algorithms over the training episodes ensures that the two regimes are visited while training. 
}

We next compare the performance of \textit{prob-DDPG} and the \textit{hid-DDPG}, rolling Z-score, {{and the model-based plug-in strategy introduced above using the cumulative $P\&L_T = \sum_{s<T} r_s$ at the end of the trading day, where $T$ is the total number of observations in the test set.  We report the ${P\&L}$ trajectories over the test day in Fig.~\ref{fig:pnl_trajectories}.
}

{\paragraph{Real data results.}
\begin{figure}[ht]
    \centering
    \includegraphics[width=0.8\textwidth]{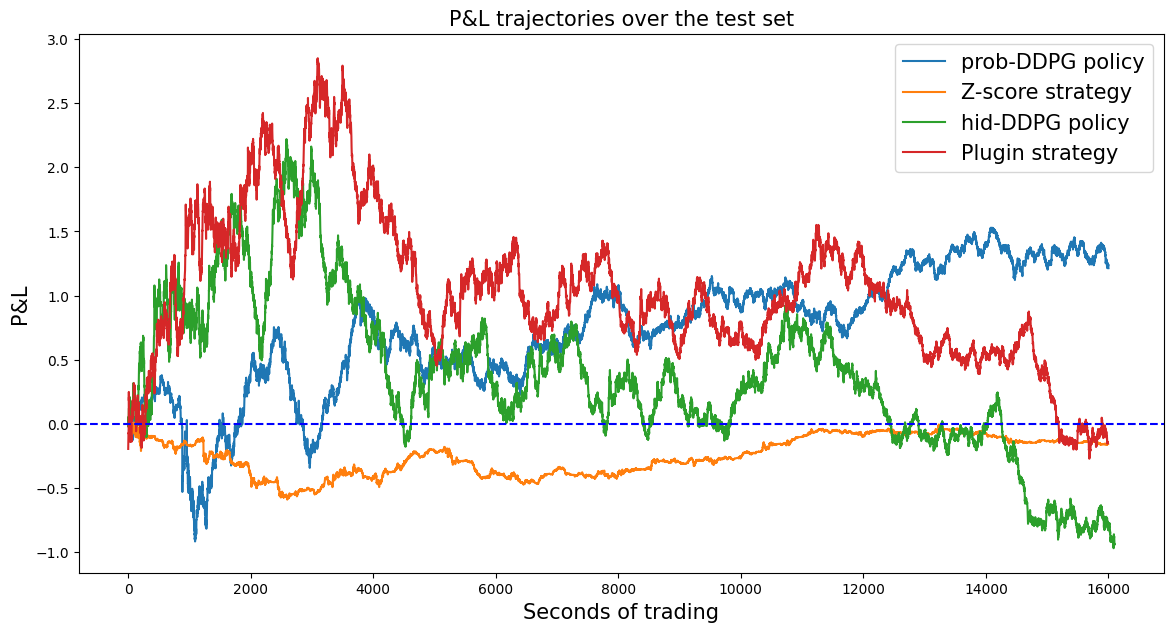}
    \caption{Comparison of realised $P\&L$ for the \textit{hid-DDPG}, \textit{prob-DDPG}, rolling Z-score and {benchmark plug-in} strategies on the normalised $\tilde{S}_t$ testing set.}
    \label{fig:pnl_trajectories}
\end{figure}

Table~\ref{tab:smh_intc_rez} reports the realised $P\&L_T$ of the four methods on the test day, after excluding the most volatile opening and closing half-hours, and Fig.~\ref{fig:pnl_trajectories} shows the corresponding cumulative trajectories. The \textit{prob-DDPG} agent achieves the highest realised cumulative $P\&L$, with a final value of $+1.296$. The other three strategies finish the evaluation window with negative cumulative $P\&L$: $-0.643$ for \textit{hid-DDPG}, $-0.146$ for the model-based plug-in rule, and $-0.170$ for the rolling $Z$-score benchmark. These performances reflect the average per-step reward, reported in basis points on the second column of Table~\ref{tab:smh_intc_rez}, showing a substantial variation over the strategies and the policies adopted, even on step-wise rewards.
\begin{table}[H]
    \centering
    
    \caption{{Final realised profit and loss $P\&L_T$ (end-of-day value of the cumulative P\&L curve) for each method on the test set (5th September), over one trading day at 1-second frequency. The last column reports the average per-step reward $r_t$ in basis points, with its standard error in parentheses. The plug-in is the model-based strategy of Section~\ref{sec:plugin}; the $Z$-score is the rolling benchmark of \cite{chan2013algorithmic}. Higher is better.}}
    \label{tab:smh_intc_rez}
    \begin{tabular}{l c c}
        \toprule
        Strategy & $P\&L_T$ & Average $r_t$ (bps) \\
        \midrule
       \textit{prob-DDPG } & $\mathbf{+1.296}$&  $\mathbf{+1.067}$ ($0.925$)\\
       \textit{hid-DDPG }  & $-0.643$         &  ${-1.069}$        ($1.494$)\\
        Plug-in            & $-0.146$         &  ${+0.025}$        ($1.629$)\\
        $Z$-score          & $-0.170$         &  ${-0.105}$        ($0.273$)\\
        \bottomrule
    \end{tabular}
\end{table}

The model-based strategy generates substantial positive $P\&L$ during the first part of the evaluation window but subsequently gives back most of these gains and ends with a slightly negative final $P\&L$. This behaviour is consistent with the structure of the model-based rule: when the estimated dynamics imply a sufficiently large expected movement relative to transaction costs, the strategy takes a persistent directional position. Its subsequent performance therefore depends strongly on whether the realised evolution of the signal remains aligned with the model-implied direction. 
A further reason for this behaviour is that the plug-in rule conditions on a single, fixed long-run level $\tilde\theta$, whereas the regime may switch during the test day. Unlike \textit{prob-DDPG}, which updates its regime posterior as new observations arrive, the plug-in cannot adapt its target level, so a directional position taken under one regime is not revised when the latent state changes.

\textit{hid-DDPG} shows a similar directional path for the cumulative $P\&L$. Over the first half of the trading day, this policy scores relatively high cumulative rewards only to lose what gained over the second half of the trading day, suggesting that the policy is strongly and inefficiently reliant on the evolution of the signal over the considered time-window. 

The simple $Z$-Score trading strategy fails to deliver positive rewards from trading consistently over the course of the trading day.

Finally, the \textit{prob-DDPG} approach does not impose the estimated mean-reverting dynamics directly on the trading decision. Instead, it conditions the policy on the posterior distribution of the latent regime, allowing the inferred information state to evolve as new observations become available. In particular, this positive performance suggests that the posterior representation of the latent regime provides useful information for the trading problem when combined with a flexible, model-agnostic control policy.

The comparison between the approaches, therefore, highlights the differences between either using an estimated parametric model directly to determine a trading rule, or providing raw information on the signal or using filtered latent-state information within a model-agnostic control framework. The last approach performs better in terms of cumulative revenues, according to these simple experiments. 
Overall, these real data results are consistent with the findings of the synthetic experiments reported above. 

We stress that these results should be interpreted as an \textit{empirical proof of concept} rather than as a comprehensive assessment of out-of-sample trading performance. The experiment is based on a single test day and is intended primarily to demonstrate that the proposed information-aware deep RL framework can be implemented on real high-frequency market data and that it can translate filtered information about latent mean-reversion dynamics into trading decisions. 

\section{Conclusions}
{
In this paper we developed and compared three DDPG-based algorithms to approximately solve an optimal trading problem where the traded signal is mean-reverting and its parameters are latent and regime-switching. Because the regime is unobservable and must be inferred from the signal, the problem is naturally cast as a partially observable Markov decision process (POMDP), in this setup, the relevant state is not the signal alone, but the belief over the latent regime. To exploit the sequential structure of the signal, each algorithm couples the DDPG actor-critic with a Gated Recurrent Unit, and the three algorithms differ in the way the latent information is represented and passed to the agent: \textit{hid-DDPG }feeds the last GRU hidden state directly to the policy;\textit{prob-DDPG} first filters the posterior probabilities of the regime and then trades on them  and \textit{reg-DDPG }forecasts the next value of the signal. 

Across a sequence of synthetic environments of increasing latent complexity, in which the long-run level, the speed of mean reversion and the volatility are governed by independent Markov chains, \textit{prob-DDPG }consistently achieves the highest rewards and the most interpretable policies. We compare the results for the simplest environment considered with those of an oracle that observes the true regime and sets an upper bound that no feasible method attains, and a na\"ive agent that ignores the regime sets a lower one. What distinguishes \textit{prob-DDPG }is that it conditions on the regime posterior which, under the POMDP formulation, is a sufficient statistic for the latent state, forecasting the next signal value (\textit{reg-DDPG}) or relying on an unstructured hidden state (\textit{hid-DDPG}) recovers that information less effectively. 

We then applied the methods to a real pairs-trading problem, trading a co-integrated portfolio of two NASDAQ-listed assets. 
This experiment provides an empirical proof of the \textit{prob-DDPG } outperforms the other strategies and is the only method to close the trading day with a positive $P\&L$. The plug-in strategy  either does not trade or takes a directional position. The regime-aware agent, by contrast, extracts a tradable signal by filtering the latent state and converts it into an actively traded, positive return. 
That a belief-based representation (\textit{prob-DDPG}) outperforms others illustrates that the structure of information, not only its availability, drives performance.
Our scenarios illustrate that an \textit{approximate} policy that uses beliefs on latent states outperforms an exact policy on a misspecified environment.

Taken together, our results show that under partial observability, the representation of information supplied to the RL agent is critical to the learned policy. Forecasting the next signal value provides limited benefit, whereas filtering the latent regime and supplying its posterior to the agent yields the best performance among the approaches considered, on both synthetic and real data. This finding is consistent with the POMDP formulation, in which the belief state is a sufficient statistic for decision-making. A natural extension is to study how such information affects learning, interaction, and equilibrium behaviour in multi-agent settings, particularly in the presence of price impact and heterogeneous information.}

\printbibliography
\newpage

\begin{appendices}
\section{The GRU network}
\label{appendix:GRU}

{
The architecture, shown in Fig.~\ref{fig:one-step-architecture}, consists of
$d_l \times (W+1)$ GRU units arranged in $d_l$ layers. The first layer reads the
sub-matrix $\es \in \R^{b \times (W+1)}$ of $\f$, containing the past and present
$W+1$ observations of $S$; subsequent layers read the hidden states of the layer
below. 
For each column $Z_k \in \R^b$ of $\es$, $k \in \{0,\dots,W\}$, the first-layer GRU
computes the \textit{reset gate}, the \textit{update gate}, the \textit{candidate
hidden state} and the \textit{hidden state} as
\begin{align}
    p_k &= \sigma(\textbf{H}_p Z_k + \textbf{U}_p h_{k-1} + \tilde b_p), \\
    z_k &= \sigma(\textbf{H}_z Z_k + \textbf{U}_z h_{k-1} + \tilde b_z), \\
    \tilde{h}_k &= \sigma(\textbf{H}_h Z_k + \textbf{U}_h (p_k * h_{k-1}) + \tilde b_h), \\
    \label{eqn:gru_net}
    h_k &= (1 - z_k) * h_{k-1} + z_k * \tilde{h}_k,
\end{align}
where $*$ is the element-wise product, $h_{-1} = 0$, and $\sigma$ is the hyperbolic
tangent (\texttt{tanh}). The reset gate $p_k \in \R^{d_h}$ selectively discards or
retains past information, while the update gate $z_k \in \R^{d_h}$ controls how much
of the previous hidden state is carried forward. With $d_h$ the hidden dimension,
$\textbf{U}_p, \textbf{U}_z, \textbf{U}_h \in \R^{d_h \times d_h}$ and $\tilde b_p $ and $\tilde b_h  \in \R^{d_h}$,
while $\textbf{H}_p, \textbf{H}_z, \textbf{H}_h \in \R^{d_h \times b}$, $h_{k-1} \in \R^{d_h}$ in the first layer and
$\R^{d_h \times d_h}$ thereafter, since deeper layers replace $Z_k$ by the hidden
state of the layer below.
\\
The network output is the last-layer hidden state at $k = W$, denoted
$o_t \in \R^b$, where the subscript records the time index of the Actor-Critic
step. The matrix $\g_t \in \R^{b \times 3}$ fed to the Actor and Critic stacks $o_t$
together with the inventory and the current signal from $\f$. Note that $o_t$ is not
a forecast of the next signal value: it encodes the temporal dependence structure of
the signal process.
\\
The GRU is updated at each iteration, before the Critic $Q$ and the Actor $\pi$. Its
parameters are trained by passing the final hidden states through a linear layer with
\texttt{LeakyReLU} activation to obtain an estimate $\tilde S_{t+1}$, and minimising the
mean squared error against $S_{t+1}$.
}

\section{Network and Model Parameters}
\label{sec:Params}

\begin{table}[H]
    \centering
    \caption{{Parameters common to all methods and experiments: the DDPG training
    setup, the simulated market, and the optimiser of the GRU filtering step.}}
    \label{tab:shared_pmts}
    
    \begin{tabular}{@{}llll@{}}
        \toprule
        \multicolumn{4}{@{}l}{\textit{DDPG training}}\\
        \midrule
        Learning rate           & $0.001$   & Train / test episodes     & $10{,}000\ /\ 500$ \\
        Batch size              & $512$     & Penalty $\lambda$   & $0.05$ \\
        Action bounds           & $[-10,10]$&  Optimiser & W-ADAM \\
        $\mu_{\text{inv}}$        & $1$       &  $\gamma$         & $0.999$ \\
        \midrule
        \multicolumn{4}{@{}l}{\textit{Simulated market}}\\
        \midrule
        Time step $\Delta t$    & $0.2$     & Steps per testing episode $n$     & $2{,}000$ \\
        $S_0$ & $1.0\$$ & $I_0$ & $0$ units\\
        \midrule
        \multicolumn{4}{@{}l}{\textit{GRU filtering step}}\\
        \midrule
        Optimiser               & W-ADAM    & Learning rate             & $0.001$ \\
        \bottomrule
    \end{tabular}
\end{table}
\begin{table}[H]
    \centering
    \caption{{The three experiments. Each row lists which parameters are modelled as
    independent Markov chains and the values they take; later experiments make more of the market latent, and hence harder the trading problem.}}
    \label{tab:sim_pmts}
    
    \begin{tabular}{@{}lccc@{}}
        \toprule
        Latent (Markov-chain) parameters & $\theta$ & $\kappa$ & $\sigma$ \\
        \midrule
        $\theta_t$                    & $\{0.9,1,1.1\}$ & $5$       & $0.2$ \\
        $\theta_t,\kappa_t$           & $\{0.9,1,1.1\}$ & $\{3,7\}$ & $0.2$ \\
        $\theta_t,\kappa_t,\sigma_t$  & $\{0.9,1,1.1\}$ & $\{3,7\}$ & $\{0.1,0.3\}$ \\
        \bottomrule
    \end{tabular}
\end{table}

\begin{table}[H]
    \centering
    \caption{{Network architectures across the three experiments. Each method uses
    two networks: a \emph{GRU} that filters the latent state from a window of $W$
    past signals, and a \emph{policy} network (the DDPG Actor - Critic). Entries are
    given as \textsf{layers\,$\times$\,width}, with the GRU look-back window $W$
    where applicable.}}
    \label{tab:architectures}
    
    \begin{tabular}{@{}ll ccc@{}}
        \toprule
        Method & Network & $\theta_t$ & $\theta_t,\kappa_t$ & $\theta_t,\kappa_t,\sigma_t$ \\
        \midrule
        \multirow{2}{*}{hid-DDPG}
            & GRU     & $20\times10,\ W{=}10$ & $20\times10,\ W{=}10$ & $20\times10,\ W{=}20$ \\
            & Policy  & $5\times20$          & $6\times20$          & $6\times20$ \\
        \addlinespace
        \multirow{2}{*}{prob-DDPG}
            & GRU     & $5\times5,\ W{=}10$ & $5\times5,\ W{=}10$ & $5\times20,\ W{=}20$ \\
            & Policy  & $5\times 20$          & $6\times20$          & $6\times20$ \\
        \addlinespace
        \multirow{2}{*}{reg-DDPG}
            & GRU     & $5\times5,\ W{=}50$ & $5\times5,\ W{=}50$ & $5\times5,\ W{=}50$ \\
            & Policy  & $5\times20$          & $6\times20$          & $6\times20$ \\
        \bottomrule
    \end{tabular}
\end{table}

\begin{table}[H] 
\centering 
\caption{VAR Model Coefficients at 1\% significance estimated from data.}
\label{tab:var_model} 

\begin{tabular}{|l|c|c|c|} 
\hline & \multicolumn{2}{c|}{{B}} & {C} \\ & $S_{t-1,\text{SMH}}$ & $S_{t-1,\text{INTC}}$ & \\
\hline
$S_{t,\text{SMH}}$ & 0.999$^{(***)}$ & 0.00001 & 0.021 \\
$S_{t,\text{INTC}}$ & 0.0007 & 0.996$^{(***)}$ & 0.003 \\
\hline
\end{tabular} 
\end{table}

\section{Figures}
\begin{figure}[H]
    \centering
    \begin{subfigure}{\textwidth}
        \centering
        \includegraphics[scale=0.4]{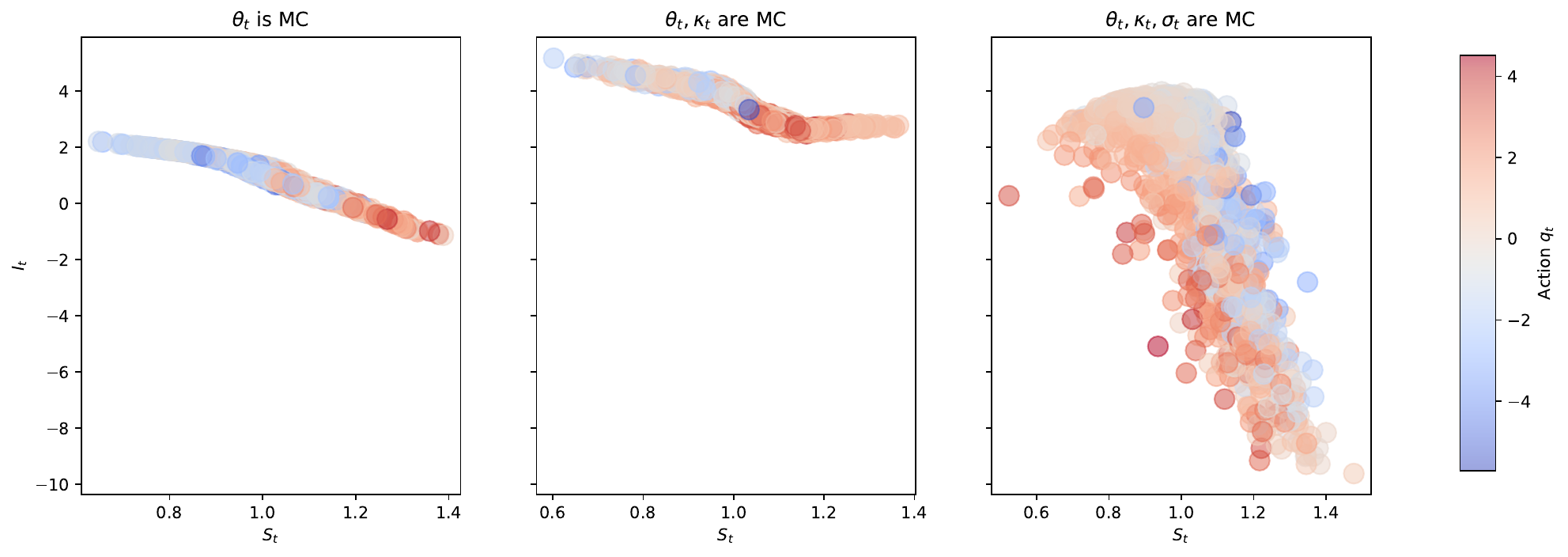}
        \caption{hid-DDPG}
        \label{fig:hid_ddpg}
    \end{subfigure}

    \begin{subfigure}{\textwidth}
        \centering
        \includegraphics[scale=0.4]{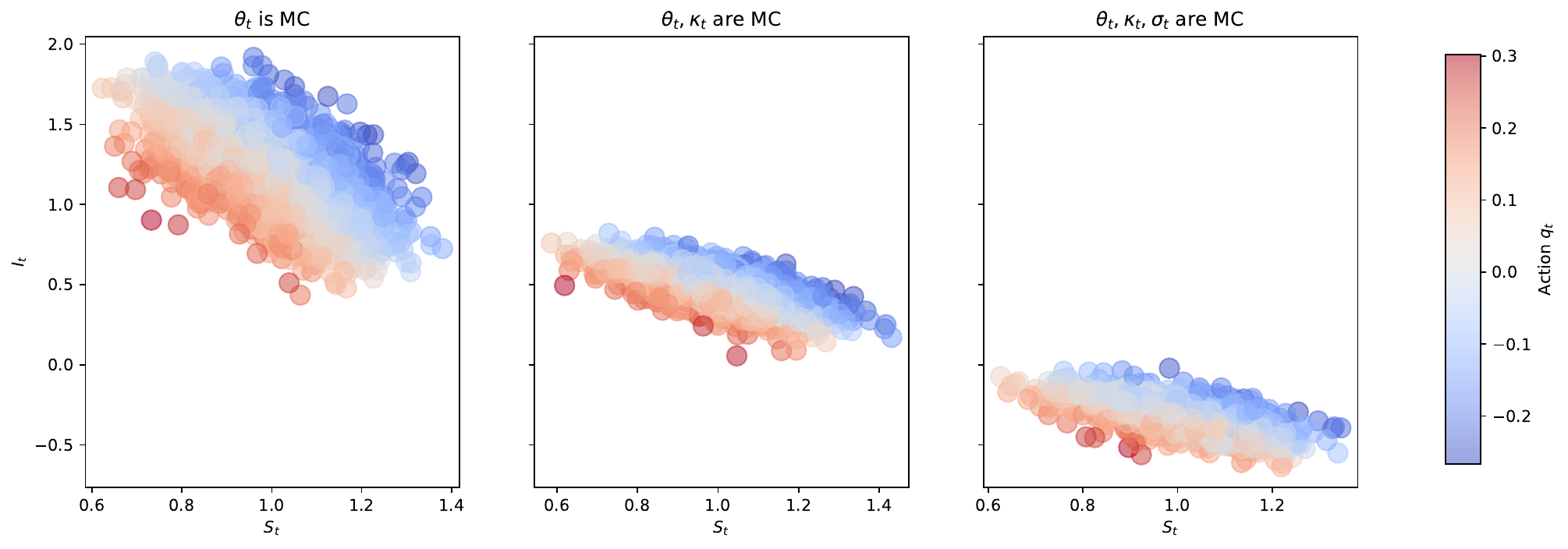}
        \caption{reg-DDPG}
        \label{fig:reg_ddpg}
    \end{subfigure}

    \begin{subfigure}{\textwidth}
        \centering
        \includegraphics[scale=0.4]{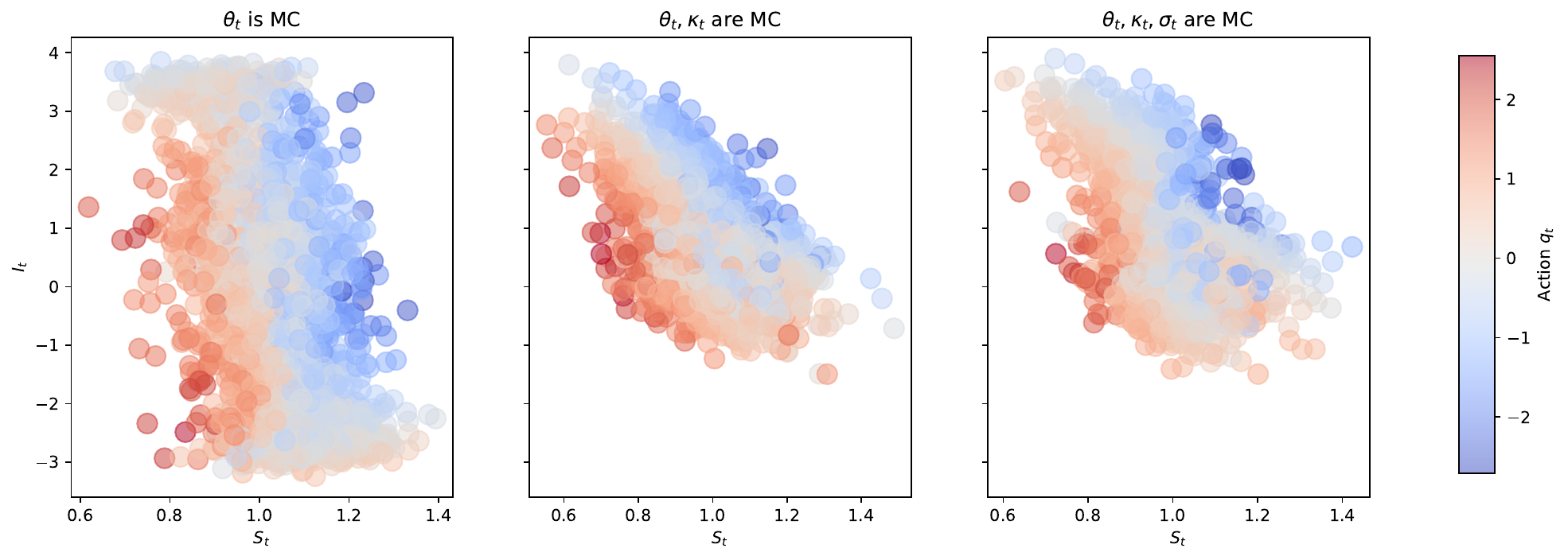}
        \caption{prob-DDPG}
        \label{fig:prob_ddpg}
    \end{subfigure}

    \caption{Magnitude of buy and sell rebalancing $q_t = I_{t+1} - I_{t}$ per level of inventory $I_t$ and signal $S_t$ for the different approaches used.}
    \label{fig:actions_varie}
\end{figure}

\begin{figure}[H]
    \centering
    \includegraphics[width=1\textwidth, angle = 270 ]{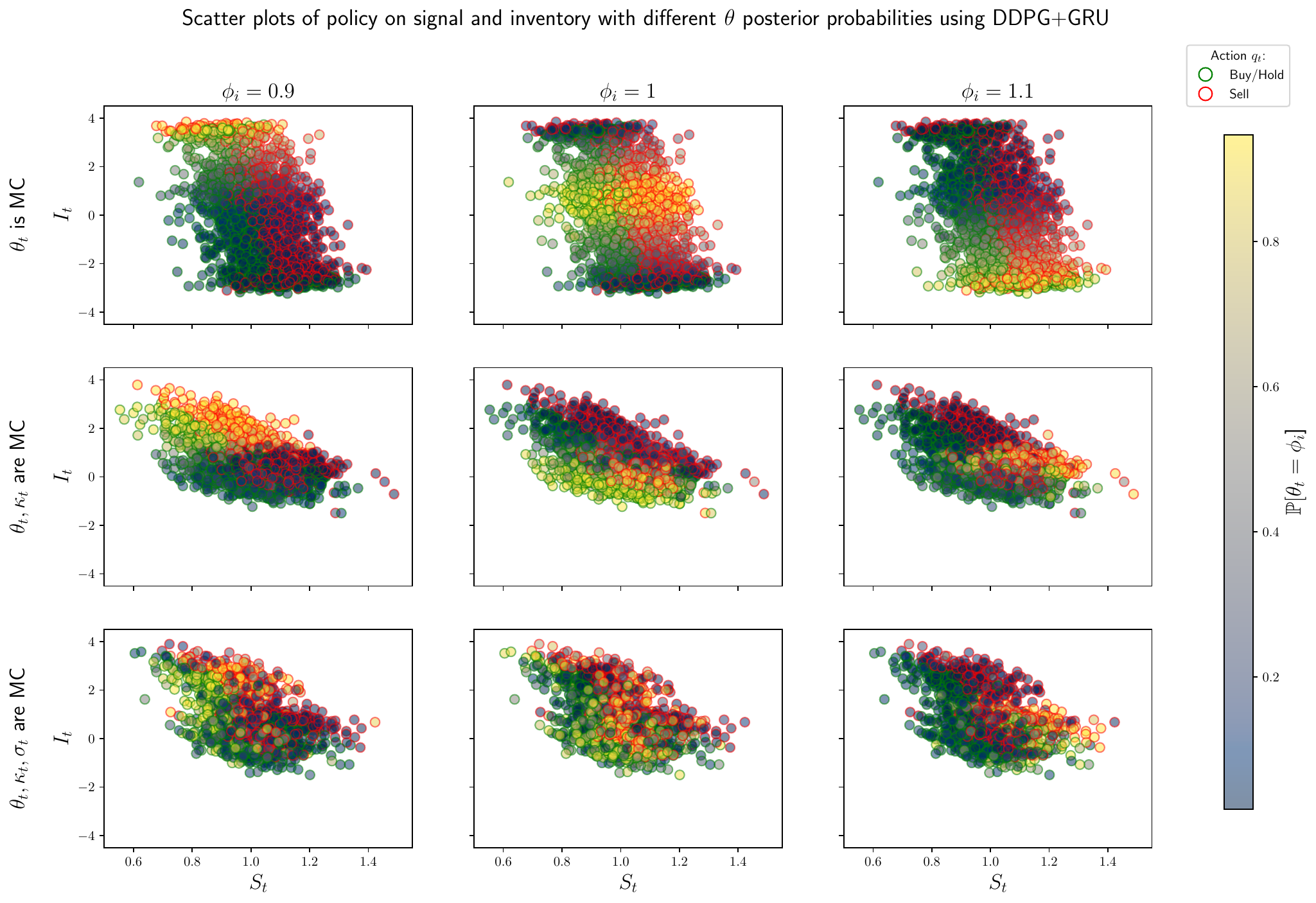}
    \caption{Policies chosen by the DDPG agent per level of inventory, price signal $S_t$ and posterior probability of mean reversion $\theta_t$ (yellow), when \textit{prob-DDPG} approach is used. The three columns correspond to the same realised states plotted against different posterior probability estimates obtained with the GRU neural network.}
    \label{fig:scatters_thetas}
\end{figure}

\newpage
\section{Algorithms}
\begin{algorithm}[H]
\caption{Deep Deterministic Policy Gradient (DDPG)}

\begin{algorithmic}[1]
\Require $W$ look-back window, $\ell$, $l$  and $N$ iteration
\State \textbf{Initialise} actor network $\pi(\mu_\pi)$ and critic network $Q(\mu_Q)$ with random weights $\mu_\pi$ and $\mu_Q$
\State \textbf{Initialise} target network and $Q_{\text{tgt}}(s,a|\mu_{Q_{\text{tgt}}})$ with weights $\mu_{Q_{\text{tgt}}} \leftarrow \mu_Q$

\For {m = 1 , \dots, N} 
  \State  \textbf{Obtain} batches of length $b$ of $S_{[0, W+2]}$ signal time series and random levels of Inventory $I$.
  \State \textbf{Initialise} $\varepsilon = 1$
  \Procedure{GRU}{} 
  \\
  \hspace{1cm} Train the GRU network if hid-DDPG;
  \EndProcedure
  \Procedure{Update Critic}{}
  \For { $\ell$}
    \State Receive initial observation batch $\f$
    \State Pass $(\{S_{u}\}^{t-W}_{u=t}$ to GRU pre-trained and obtain $\{\Phi_k\}, k=1,2,3)$ (if prob-DDPG);
    \State Pass $(\{S_{u}\}^{t-W}_{u=t}$ to GRU pre-trained and obtain $\tilde S_{t+1}$ (if reg-DDPG);
    \State Pass $(\{S_{u}\}^{t-W}_{u=t}$ to GRU to train if (hid-DDPG);
    \State obtain \(\s\);
    \State Select inventory $I = \pi(\s|\mu_\pi) + \mathcal{N}(0, \varepsilon)$ according to the current policy and noise;
      \State Execute action $I$ and observe reward $r$;  and next state $\s'$;
      \State Observe next states $\s'$;
    \State Set $y^{(i)} = r^{(i)} + \gamma Q_{\text{tgt}}(\s'^{(i)}, \pi(\s'^{(i)}|\mu_\pi)|\mu_{Q_{\text{tgt}}})$ ;

    \State $\LL_1 = \frac{1}{b} \sum_{i}^{b} (Q(\s^{(i)}, I^{(i)}|\mu_Q) - y^{(i)})^2$;

    \State Set $\mu_Q = \mu_{Q_{\text{tgt}}}$;
    
    \EndFor
  \EndProcedure
  \Procedure{Update Actor}{}
  \For l 
    \State Receive initial observation batch $\f$;
    \State Pass $(\{S_{u}\}^{t-W}_{u=t}$ to GRU pre-trained and obtain $\{\Phi_k\}, k=1,2,3)$ (if prob-DDPG);
    \State Pass $(\{S_{u}\}^{t-W}_{u=t}$ to GRU pre-trained and obtain $\tilde S_{t+1}$ (if reg-DDPG);
    \State Pass $(\{S_{u}\}^{t-W}_{u=t}$ to GRU to train if (hid-DDPG); 
    \State Obtain \(\s\);
    \State Get $I = \pi(\s|\mu_\pi)$;
    \State Minimise $\L_2 = -\frac{1}{b} \sum^b_{i =1} (Q(\s^{(i)}, \pi(\s^{(i)}|\mu_\pi)|\mu_Q)$ through gradient descent over
    $$     \nabla_{\mu_\pi}\L_2 = \frac1b\sum^b_{i=1}\left[ \nabla_{a}Q(\s^{(i)}, a^{(i)}|\mu_Q)|_{a^{(i)} = \pi(\s^{(i)}|\mu_\pi)} \nabla_{\mu_\pi}(\s^{(i)}|\mu_\pi) \right] $$
  \EndFor
  \EndProcedure
  \State Decrease $\varepsilon$;
\EndFor
\end{algorithmic}
\label{algo:ddpg}
\end{algorithm}
\end{appendices}

\end{document}

%% file: preamble.tex
\usepackage{amssymb}
\DeclareMathOperator*{\argmin}{arg\,min}
\newcommand{\E}{{\mathbb E}}
\newcommand{\F}{{\mathcal F}}
\renewcommand{\P}{{\mathbb P}}
\renewcommand{\L}{\mathcal L}
\newcommand{\R}{\mathbb R}
\newcommand{\f}{\mathbf F}
\newcommand{\s}{\mathbf G}
\newcommand{\es}{\mathbf Z}
\newcommand{\g}{\mathbf G}

\newcommand{\wPhi}{{\widehat{\Phi}}}

\newcommand{\LL}{\mathcal{L}}
\newcommand{\mcA}{\mathcal{A}}
\newcommand{\mcI}{\mathcal{I}}
\newcommand{\K}{{\mathcal{K}}}

\newcommand{\dW}{{\rm{d} W}}
\newcommand{\dS}{{\rm{d} S}}
\newcommand{\dt}{{\rm{d} t}}
\newcommand{\du}{{\rm{d} u}}

\tikzstyle{node}=[shape=circle,draw=blue,fill=blue!10]
\tikzstyle{a}=[shape=circle,draw=green,fill=green!10]
\tikzstyle{q}=[shape=circle,draw=purple,fill=purple!10]
\tikzstyle{pi}=[shape=circle,draw=purple,fill=purple!10]
\tikzstyle{p}=[shape=circle,draw=purple,fill=yellow!10]
\tikzstyle{hidden}=[shape=circle,draw=red,fill=yellow!30]
\tikzstyle{gru}=[shape=rectangle,draw=black!50,fill=lime!20]
\tikzstyle{ffn}=[shape=rectangle,draw=black!50,fill=lime!20]

%% file: onestep.tex
\begin{figure}[H]
    \centering
    \begin{tikzpicture}[scale=0.6,every node/.style={transform shape},minimum width=1.0cm]
    
    \node[hidden] (h0) at (-2,0) {$0$};


    \node[node] (y1) at (0,-1.5) {$S_{t-W}$};
    \node[gru] (gru1) at (0,0) {$~GRU~$};
    \node[gru] (gru11) at (0,1.5) {$~GRU~$};
    \node[hidden] (h1) at (0,3) {$h_{0}$};
    
    \draw [->] (h0) to (gru1);
    \draw [->] (y1) to (gru1);
    \draw [->] (gru1) to (gru11);
    \draw [->] (gru11) to (h1);

    
    \node[node] (y2) at (2,-1.5) {$S_{t-W+1}$};
    \node[gru] (gru2) at (2,0) {$~GRU~$};
    \node[gru] (gru21) at (2,1.5) {$~GRU~$};
    \node[hidden] (h2) at (2,3) {$h_{1}$};

    \draw [->] (y2) to (gru2);
    \draw [->] (gru1) to (gru2);
    \draw [->] (gru2) to (gru21);
    \draw [->] (gru11) to (gru21);    
    \draw [->] (gru21) to (h2);

    
    \node[node] (y3) at (4,-1.5) {$S_{t-W+2}$};
    \node[gru] (gru3) at (4,0) {$~GRU~$};
    \node[hidden] (h3) at (4,3) {$h_{2}$};
    \node[gru] (gru31) at (4,1.5) {$~GRU~$};    
    
    \draw [->] (y3) to (gru3);
    \draw [->] (gru2) to (gru3);
    \draw [->] (gru3) to (gru31);
    \draw [->] (gru21) to (gru31);    
    \draw [->] (gru31) to (h3);


    \node[node] (y4) at (7,-1.5) {$S_{t}$};

    \node[gru] (gru4) at (7,0) {$~GRU~$};
    \node[gru] (gru41) at (7,1.5) {$~GRU~$};
    \node[hidden] (h4) at (7,3) {$h_W$};
    
    \draw [->,dashed] (gru3) to (gru4);
    \draw [->] (y4) to (gru4);
    \draw [->,dashed] (gru31) to (gru41);
    \draw [->] (gru4) to (gru41);
    \draw [->] (gru41) to (h4);

    
    \node[ffn] (fq) at (13,4.5) {$~Q-FFN~$};
    \node[ffn] (fpi) at (10.5,1.5) {$~\pi-FFN~$};
    \draw [->] (h4) to [out=0, in=180] (fq);
    \draw [->] (h4) to [out=0, in=90] (fpi);



    \node[a] (a) at (12,3) {$~I_{t+1}~$};
    \node[q] (q) at (15,3) {$~Q~$};
    \node[node] (y6) at (10,-1.6) {$I_{t}$};
    \draw [->] (fq) to [out=0,in=90] (q);    
    \draw [->] (a) to [out=90,in=-135] (fq);
    \draw [->] (fpi) to [out=0,in=-90] (a);
    \draw [->] (y6)  to (fpi);
    \draw [->] (y6)  to [out=0, in=-90] (fq);    


    \draw[->] (y4) to [out=0,in=-135] (fpi);
    \draw[->] (y4) to [out=45,in=180] (fq);    
    
    \end{tikzpicture}
    \caption{Directed graph representation of the neural network architecture for the one-step architecture. In the figure $d_\ell=2$. }
    \label{fig:one-step-architecture}
\end{figure}

%% file: twostep.tex
\begin{figure}[H]
    \centering
    \begin{tikzpicture}[scale=0.6,every node/.style={transform shape},minimum width=1.0cm]
    
    \node[hidden] (h0) at (-2,0) {$0$};


    \node[node] (y1) at (0,-1.5) {$S_{t-W}$};
    \node[gru] (gru1) at (0,0) {$~GRU~$};
    \node[gru] (gru11) at (0,1.5) {$~GRU~$};
    \node[hidden] (h1) at (0,3) {$h_{1}$};
    
    \draw [->] (h0) to (gru1);
    \draw [->] (y1) to (gru1);
    \draw [->] (gru1) to (gru11);
    \draw [->] (gru11) to (h1);

    
    \node[node] (y2) at (2,-1.5) {$S_{t-W+1}$};
    \node[gru] (gru2) at (2,0) {$~GRU~$};
    \node[gru] (gru21) at (2,1.5) {$~GRU~$};
    \node[hidden] (h2) at (2,3) {$h_{2}$};

    \draw [->] (y2) to (gru2);
    \draw [->] (gru1) to (gru2);
    \draw [->] (gru2) to (gru21);
    \draw [->] (gru11) to (gru21);    
    \draw [->] (gru21) to (h2);

    
    \node[node] (y3) at (4,-1.5) {$S_{t-W+2}$};
    \node[gru] (gru3) at (4,0) {$~GRU~$};
    \node[hidden] (h3) at (4,3) {$h_{3}$};
    \node[gru] (gru31) at (4,1.5) {$~GRU~$};    
    
    \draw [->] (y3) to (gru3);
    \draw [->] (gru2) to (gru3);
    \draw [->] (gru3) to (gru31);
    \draw [->] (gru21) to (gru31);    
    \draw [->] (gru31) to (h3);


    \node[node] (y4) at (7,-1.5) {$S_{t}$};
    \node[node] (y5) at (7,-3) {$I_{t}$};
    \node[gru] (gru4) at (7,0) {$~GRU~$};
    \node[gru] (gru41) at (7,1.5) {$~GRU~$};
    \node[hidden] (h4) at (7,3) {$h_W$};
    
    \draw [->,dashed] (gru3) to (gru4);
    \draw [->] (y4) to (gru4);
    \draw [->,dashed] (gru31) to (gru41);
    \draw [->] (gru4) to (gru41);
    \draw [->] (gru41) to (h4);

    
    \node[ffn] (f) at (12,3) {$~g-FFN~$};
    \node[p] (p1) at (10,1.5) {$~p_1~$};
    \node[p] (p2) at (12,1.5) {$~p_2~$};
    \node[p] (p3) at (14,1.5) {$~p_3~$};
    
    \draw [->] (h4) to (f);
    \draw [->] (f) to (p1);
    \draw [->] (f) to (p2);
    \draw [->] (f) to (p3);
    
    \node[q] (q) at (12,-1) {$~Q~$};
    \node[a] (a) at (14,-3) {$~I_{t+1}~$};
    \node[ffn] (fq) at (10,0) {$~Q-FFN~$};
    \node[ffn] (fpi) at (14,0) {$~\pi-FFN~$};

    \draw[->] (p1) to  (fq);
    \draw[->] (p2) to [out=225,in=45] (fq);
    \draw[->] (p3) to [out=225,in=0] (fq);
    \draw[->] (a) to [out=180,in=-90] (fq);
    \draw[->] (y4) to [out=0,in=180] (fq);
    \draw[->] (fq) to (q);

    \draw[->] (p1) to [out=-45,in=180] (fpi);
    \draw[->] (p2) to [out=-45,in=135] (fpi);
    \draw[->] (p3) to (fpi); 
    \draw[->] (y4) to [out=-45,in=-90] (fpi);
    \draw[->] (y5) [out=0,in=-125] to (fpi); 
    \draw[->] (y5) to [out=45, in=-120] (fq);

    \draw[->]  (fpi) to [out=-45,in=90] (a); 

    \end{tikzpicture}
    \caption{Directed graph representation of the neural network architecture for the two-step architecture with regime filtering.}
    \label{fig:two-step-architecture}
\end{figure}

%% file: twostep_reg.tex
\begin{figure}[H]
    \centering
    \begin{tikzpicture}[scale=0.6,every node/.style={transform shape},minimum width=1.0cm]
    
    \node[hidden] (h0) at (-2,0) {$0$};


    \node[node] (y1) at (0,-1.5) {$S_{t-W}$};
    \node[gru] (gru1) at (0,0) {$~GRU~$};
    \node[gru] (gru11) at (0,1.5) {$~GRU~$};
    \node[hidden] (h1) at (0,3) {$h_{1}$};
    
    \draw [->] (h0) to (gru1);
    \draw [->] (y1) to (gru1);
    \draw [->] (gru1) to (gru11);
    \draw [->] (gru11) to (h1);

    
    \node[node] (y2) at (2,-1.5) {$S_{t-W+1}$};
    \node[gru] (gru2) at (2,0) {$~GRU~$};
    \node[gru] (gru21) at (2,1.5) {$~GRU~$};
    \node[hidden] (h2) at (2,3) {$h_{2}$};

    \draw [->] (y2) to (gru2);
    \draw [->] (gru1) to (gru2);
    \draw [->] (gru2) to (gru21);
    \draw [->] (gru11) to (gru21);    
    \draw [->] (gru21) to (h2);

    
    \node[node] (y3) at (4,-1.5) {$S_{t-W+2}$};
    \node[gru] (gru3) at (4,0) {$~GRU~$};
    \node[hidden] (h3) at (4,3) {$h_{3}$};
    \node[gru] (gru31) at (4,1.5) {$~GRU~$};    
    
    \draw [->] (y3) to (gru3);
    \draw [->] (gru2) to (gru3);
    \draw [->] (gru3) to (gru31);
    \draw [->] (gru21) to (gru31);    
    \draw [->] (gru31) to (h3);


    \node[node] (y4) at (7,-1.5) {$S_{t}$};
    \node[node] (y5) at (7,-3) {$I_{t}$};
    \node[gru] (gru4) at (7,0) {$~GRU~$};
    \node[gru] (gru41) at (7,1.5) {$~GRU~$};
    \node[hidden] (h4) at (7,3) {$h_W$};
    
    \draw [->,dashed] (gru3) to (gru4);
    \draw [->] (y4) to (gru4);
    \draw [->,dashed] (gru31) to (gru41);
    \draw [->] (gru4) to (gru41);
    \draw [->] (gru41) to (h4);

    
    \node[ffn] (f) at (12,3) {$~\tilde{S}_{t+1}~$};
        
    \draw [->] (h4) to (f);

    \node[q] (q) at (12,-1) {$~Q~$};
    \node[a] (a) at (14,-3) {$~I_{t+1}~$};
    \node[ffn] (fq) at (10,0) {$~Q-FFN~$};
    \node[ffn] (fpi) at (14,0) {$~\pi-FFN~$};

    \draw[->] (f) to  [out=-135,in=90] (fq);
    \draw[->] (a) to [out=180,in=-90] (fq);
    \draw[->] (y4) to [out=0,in=180] (fq);
    \draw[->] (fq) to (q);

    \draw[->] (f) to [out=-45,in=90] (fpi);
    \draw[->] (y4) to [out=-45,in=-90] (fpi);
    \draw[->] (y5) [out=0,in=-125] to (fpi); 
    \draw[->] (y5) to [out=45, in=-120] (fq);

    \draw[->]  (fpi) to [out=-45,in=90] (a); 

    \end{tikzpicture}
    \caption{Directed graph representation of the neural network architecture for the two-step architecture with price prediction.}
    \label{fig:two-step-architecture2}
\end{figure}